\documentclass[fleqn,usenatbib]{mnras}

\usepackage{newtxtext,newtxmath}

\usepackage[T1, T2A]{fontenc}
\usepackage{ae,aecompl}

\usepackage{graphicx}	
\usepackage{amsmath}	
\usepackage{amssymb}	

\usepackage{multirow}
\usepackage{adjustbox}

\usepackage[verbose]{placeins}
\usepackage{blindtext}

\title[Rest-frame UV Properties of luminous LAEs]{Rest-frame UV Properties of Luminous Strong Gravitationally Lensed Ly$\alpha$ Emitters from the BELLS GALLERY Survey}

\author[R. Marques-Chaves et al.]{
R. Marques-Chaves,$^{1,2,3}$
I. P\'{e}rez-Fournon,$^{2,3}$
Y. Shu,$^{4}$
L. Colina,$^{1,5}$
A. Bolton,$^{6}$
\newauthor
J. \'{A}lvarez-M\'{a}rquez,$^{1}$
J. Brownstein,$^{7}$
M. Cornachione,$^{8}$
S. Geier,$^{9,2}$
C. Jim\'{e}nez-\'{A}ngel,$^{2,3}$
\newauthor
T. Kojima,$^{10,11}$
S. Mao,$^{12,13}$
A. Montero-Dorta,$^{14}$
M. Oguri,$^{11,15}$
M. Ouchi,$^{10,15}$
\newauthor
F. Poidevin,$^{2,3}$
R. Shirley,$^{2,3}$ and 
Z. Zheng$^{7}$
\\
\\
$^{1}$Centro de Astrobiolog\'ia (CSIC-INTA), Carretera de Ajalvir, 28850 Torrej\'on de Ardoz, Madrid, Spain. E-mail: rmarques@cab.inta-csic.es \\
$^{2}$Instituto de Astrof\'\i sica de Canarias, C/V\'\i a L\'actea, s/n, E-38205 San Crist\'obal de La Laguna, Tenerife, Spain\\
$^{3}$Universidad de La Laguna, Dpto. Astrof\'\i sica, E-38206 San Crist\'obal de La Laguna, Tenerife, Spain\\
$^{4}$Institute of Astronomy, University of Cambridge, Madingley Road, Cambridge CB3 0HA, UK\\
$^{5}$Cosmic Dawn Center (DAWN)\\
$^{6}$NSF's National Optical-Infrared Astronomy Research Laboratory, 950 North Cherry Ave, Tucson, AZ 85719, USA\\
$^{7}$Department of Physics and Astronomy, University of Utah, 115 S. 1400 E., Salt Lake City, UT 84112, USA\\
$^{8}$Department of Physics, United States Naval Academy, 572C Holloway Road, Annapolis, MD 21402, USA\\
$^{9}$GRANTECAN, Cuesta de San Jos\'{e} s/n, E-38712, Bre\~{n}a Baja, La Palma, Spain\\
$^{10}$Institute for Cosmic Ray Research, The University of Tokyo, 5-1-5 Kashiwanoha, Kashiwa, Chiba 277-8582, Japan\\
$^{11}$Department of Physics, The University of Tokyo, 7-3-1 Hongo, Bunkyo-ku, Tokyo 113-0033, Japan\\
$^{12}$Physics Department and Tsinghua Center for Astrophysics, Tsinghua University, Beijing, 100084, China\\
$^{13}$National Astronomical Observatories, Chinese Academy of Sciences, 20 Datun Road, Beijing, 100101, China\\
$^{14}$Departamento de F\'isica Matem\'atica, Instituto de F\'isica, Universidade de S\~ao Paulo, Rua do Mat\~ao 1371, CEP 05508-090, S\~ao Paulo, Brazil\\
$^{15}$Kavli Institute for the Physics and Mathematics of the Universe, University of Tokyo, Kashiwa, Chiba 277-8583, Japan\\
}

\date{Accepted: 2019 December 4. Received: 2019 December 4; in original form: 2019 November 5.}

\pubyear{2019}

\begin{document}
\label{firstpage}
\pagerange{\pageref{firstpage}--\pageref{lastpage}}
\maketitle

\begin{abstract}
We present deep rest-frame UV spectroscopic observations using the Gran Telescopio Canarias of six gravitationally lensed Ly$\alpha$ emitters (LAEs) at $2.36 < z < 2.82$ selected from the BELLS GALLERY survey. By taking the magnifications into account, we show that LAEs can be as luminous as $L_{\rm Ly\alpha} \simeq 30 \times 10^{42}$~erg~s$^{-1}$ and $M_{\rm UV} \simeq -23 $~(AB) without invoking an AGN component, in contrast with previous findings.
We measure Ly$\alpha$ rest-frame equivalent widths, $EW_{0}~\rm (Ly\alpha)$, ranging from 16\AA{ }to 50\AA{ }and Ly$\alpha$ escape fractions, $f_{\rm esc}~ \rm (Ly\alpha)$, from 10\% to 40\%. Large $EW_{0}~ \rm (Ly\alpha)$ and $f_{\rm esc}~ \rm (Ly\alpha)$ are found predominantly in LAEs showing weak low-ionization ISM absorption  ($EW_{0} \lesssim 1$\AA) and narrow Ly$\alpha$ profiles ($\lesssim 300$~km~s$^{-1}$ FWHM) with their peak close ($\lesssim 80$~km~s$^{-1}$) to their systemic redshifts, suggestive of less scatter from low H~{\sc i} column densities that favours the escape of Ly$\alpha$ photons. 
We infer stellar metallicities of $Z/Z_{\odot} \simeq 0.2$ in almost all LAEs by comparing the P-Cygni profiles of the wind lines N~{\sc v}1240\AA{ }and C~{\sc iv}1549\AA{ }with those from stellar synthesis models. 
We also find a trend between $M_{\rm UV}$ and the velocity offset of ISM absorption lines, such as the most luminous LAEs experience stronger outflows. 
The most luminous LAEs show star formation rates up to $\simeq 180$ $M_{\odot}$ yr$^{-1}$, yet they appear relatively blue ($\beta_{\rm UV} \simeq -1.8$ to $-2.0$) showing evidence of little dust attenuation ($E(B-V) = 0.10-0.14$). These luminous LAEs may be particular cases of young starburst galaxies that have had no time to form large amounts of dust. 
If so, they are ideal laboratories to study the early phase of massive star formation, stellar and dust mass growth, and chemical enrichment histories of starburst galaxies at high-$z$. 

\end{abstract}




\section{Introduction}

Ly$\alpha$ emitters (LAEs) are young star-forming galaxies that emit Ly$\alpha$ radiation from the photoionization of neutral hydrogen by young hot stars. LAEs have been detected mainly with narrow-band imaging of their Ly$\alpha$ emission line \citep[typically defined by rest-frame equivalent widths $EW_{0} > 20$ \AA; e.g.,][]{ajiki2003, ouchi2005, ouchi2008}. 

Multi-wavelength follow-up imaging and spectroscopy have revealed that typical ($L^{*}$) LAEs have relatively low stellar masses \citep[$M_{*} \simeq 10^{7} - 10^{9}$ $M_{\odot}$; e.g.,][]{gawiser2007, ono2010b, ono2010a, guaita2011, kusakabe2018}, show modest star-formation rates \citep[$\rm SFR \sim 10$ $M_{\odot}$ yr$^{-1}$, e.g.,][]{nakajima2012, sobral2018b}, and present, on average, low gas-phase metallicity \citep[e.g.,][]{finkelstein2011a, nakajima2012, nakajima2013, amorin2017, kojima2017}. 
They show very compact and clumpy morphologies (effective radius $r_{\rm eff} \simeq 1$ kpc) with little to no evolution with redshift \citep[e.g.,][]{bond2009, taniguchi2009, bond2012, hernan2017, cornachione2018, paulino2018, shibuya2018b}. 

Since LAEs have low stellar mass and present strong Ly$\alpha$ emission, which is indicative of a young starburst, they have been used as probes of the very first galaxies, holding important clues to the formation and evolution of galaxies at the time when the Universe was still young \citep[e.g.,][]{finkelstein2013, bouwens2015, zitrin2015, laporte2017, hashimoto2018, matthee2018, hashimoto2019}.

Although LAEs may share several of its physical properties with the Lyman break galaxy (LBG) population \citep[e.g.,][]{shapley2003, verhamme2008, schaerer2011, du2018, santos2019}, in particular those showing strong Ly$\alpha$ emission, narrow-band selected LAEs are typically fainter in the ultra-violet (UV) continuum than LBGs, given the nature of the search strategy.
Because of that, the characterization of the physical properties of LAEs has been possible using mainly stacking techniques over large samples of individual spectra or images \citep[e.g.,][]{ono2010a, wardlow2014, momose2016, nakajima2018}. 

Another way to study in detail these high-redshift galaxies is to use the natural magnification and amplification produced by gravitational lensing.
The Baryon Oscillation Spectroscopic Survey Emission-Line Lens Survey for the GALaxy-Ly$\alpha$ EmitteR sYstems \citep[BELLS GALLERY:][]{shu2016a} led to the discovery of 187 galaxy-scale strong gravitational lens candidates with LAEs at $2<z<3$ as background sources. Among these, 21 of the highest quality candidates were observed with {\it Hubble Space Telescope} ({\it HST}) confirming the lensing nature \citep{shu2016b} and providing a detailed characterization of their rest-frame UV continuum surface brightness profiles and substructure down to scales of 100~pc \citep{cornachione2018, ritondale2019}.

The BELLS GALLERY project presented in \cite{shu2016a} is the first and still unique dedicated survey aimed to search for strong gravitational lensed systems with LAEs as background sources. By taking advantage of the boost on the observable flux and the improved spatial resolution provided by gravitational lensing, many aspects of the physical properties of LAEs can be studied in much more detail than in non-lensed systems.
In this paper we present optical spectroscopic and imaging follow-up observations with the Gran Telescopio Canarias (GTC) and the William Herschel Telescope (WHT) of six gravitationally lensed LAEs from the BELLS GALLERY survey. 

The paper is structured as follows. In Section \ref{sec:data} we describe our sample and the spectroscopic and imaging observations. In Section \ref{sec: lens} we present the lens model of BG1501+3042, a new confirmed lensed LAE.  The analysis of imaging and spectroscopic data are described in Sections \ref{sec:image} and \ref{sec:spec}, respectively. Finally, in Sections \ref{sec:discussion} and \ref{sec:conclusion} we discuss our results and summarize our main findings. Throughout this work, we adopted a concordance cosmology with $\Omega_{\rm m} = 0.274$, $\Omega_{\Lambda} = 0.726$, and $H_{0} = 70$ km s$^{-1}$ Mpc$^{-1}$. All magnitudes are given in the AB system.

\section{Sample and Observations}\label{sec:data}

\subsection{Sample Selection}\label{sample_selection}

From the \textit{HST} sample analyzed in \cite{shu2016b}, we restricted the GTC spectroscopic follow-up observations to four LAEs showing large $HST$ $V$-band flux densities and large image separations. This allows us to obtain high signal-to-noise ratio (S/N) spectra, avoiding at the same time large flux contamination from the foreground lens galaxies. 
These are SDSS J020121.39+322829.6, SDSS J074249.68+334148.9, SDSS J075523.52+344539.5, and SDSS J091859.21+510452.5 (hereafter BG0201+3228, BG0742+3341, BG0755+3445, and BG0918+5104, where ``BG'' stands for BELLS GALLERY). 

\begin{figure*}
  \centering
  \includegraphics[width=0.75\textwidth]{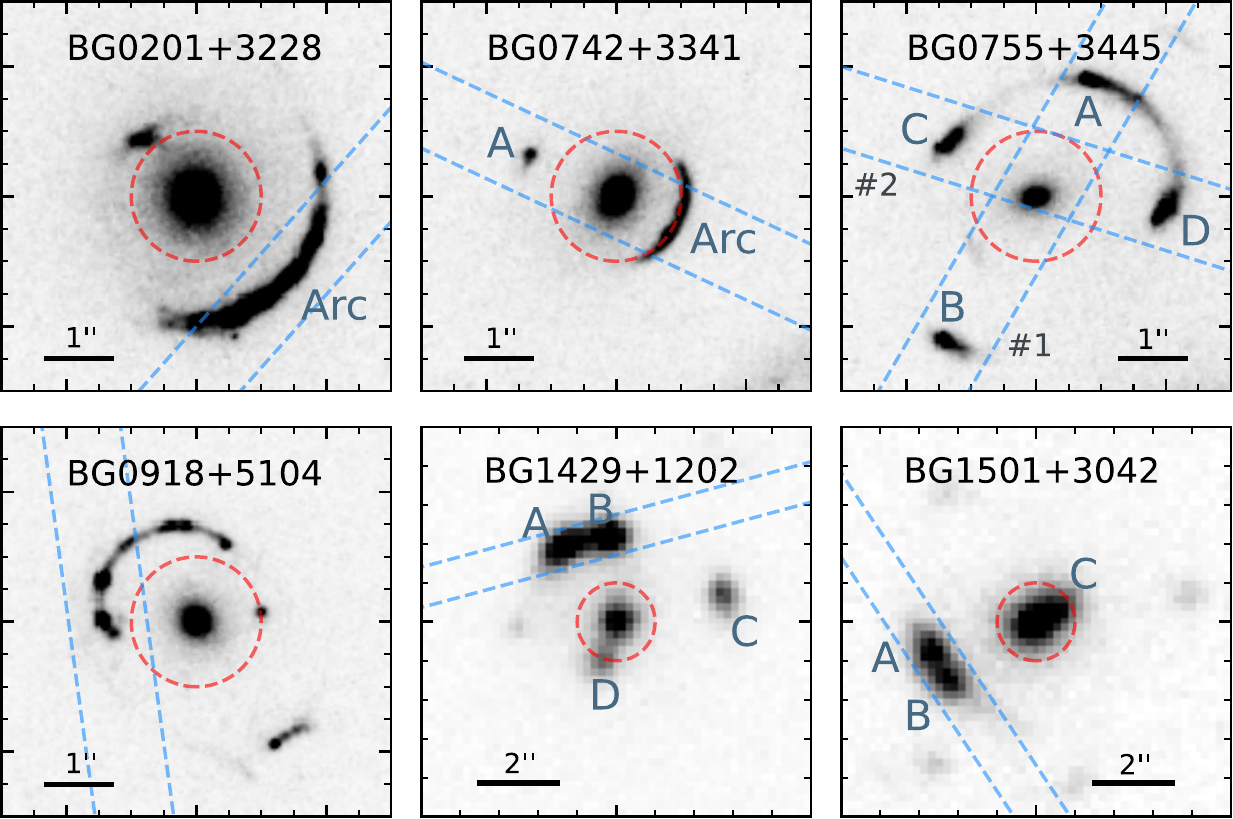}
  \caption{Cutouts of the 6 lensed LAEs analysed in this work from F606W images obtained by \textit{HST} (BG0201+3228, BG0742+3341, BG0755+3445, and BG0918+5104) and GTC (BG1429+1202 and BG1501+3042). The orientations of the GTC/OSIRIS long slit are marked with blue dashed lines, as well as the position of the spectroscopic 1$^{\prime \prime}$-radius BOSS fiber (red dashed circles). The images are centered on the LRGs and oriented such that north is up and east is to the left.}
  \label{fig:2rgb}
\end{figure*}

The LAEs BG0201+3228, BG0742+3341, and BG0918+5104 are magnified by  factors of $\mu = 15\pm3$, $16\pm3$, and $18\pm3$, respectively \citep[][]{shu2016b}, and show similar lensed morphologies, with bright extended arcs and  relatively compact counter-images on the other side of the lens in a {\it cusp} configuration. 
BG0755+3445 shows an Einstein-cross-like configuration composed of four bright lensed images in the image plane (A to D in Figure \ref{fig:2rgb}), all of them observed with the GTC, and is magnified by $\mu = 14\pm2$.

We also re-observed BG1429+1202, discussed already in \cite{marques2017} \citep[see also][]{rigby2017a, chisholm2019}, this time with higher spectral resolution and deeper imaging. This system comprises four images in the observer plane with a total magnification of $8.8$ \citep[][]{marques2017}, forming a {\it fold} configuration: a bright lensed image pair, A and B (with a separation of $\approx 1.5^{\prime \prime}$), and two fainter images, C and D (see Figure \ref{fig:2rgb}). The GTC long slit was positioned so as to encompass the two brightest lensed images A and B. 

In addition, we observed another promising lensed system candidate from the parent sample composed of 166 BELLS GALLERY of \cite{shu2016a}. SDSS J150114.60+304230.8, hereafter BG1501+3042, shows bluish (i.e. $r-i \simeq 0$) features $\approx 3.7^{\prime \prime}$ SE from the luminous red galaxy (LRG) at $z_{\rm lens} = 0.638$, and its BOSS spectrum (Plate-MJD-Fiber: 3875-55364-935; \citealt{blanton2017}) shows an emission line at $\simeq 4438$ \AA, likely Ly$\alpha$ emission at $z \simeq 2.65$.

\subsection{Spectroscopic data}\label{reduction}

The spectroscopic survey was conducted with the Optical System for Imaging and low-Intermediate Resolution Integrated Spectroscopy instrument (OSIRIS)\footnote{\url{http://www.gtc.iac.es/instruments/osiris/}} mounted at the GTC, at the Observatorio del Roque de los Muchachos. 
The observations were obtained in service mode over ten different nights, between 2017 January 26 and 2018 February 16, as part of the GTC programs GTC47-16B, GTC67-17A, and GTCMULTIPLE2F-17B (PI: R.~Marques-Chaves). 
The seeing conditions ranged between $0.7^{\prime \prime}$ and $1.1^{\prime \prime}$. 
We used $1.2^{\prime \prime}$ and $1.0^{\prime \prime}$-wide long slits oriented to cover the regions of maximum emission in order to maximize the S/N (see Figure \ref{fig:2rgb}). 
Depending on each LAE (details of these observations are listed in Table \ref{tab2}), the total integration time ranged between 2100 and 7200~s. The spectra were obtained with the R2500V, R2500R, R2000B, and R1000B grisms, providing a spectral coverage of 3630-7685 \AA. The data were reduced with standard {\sc Iraf} tasks. 1D spectra were extracted and corrected for the instrumental response using spectroscopic observations of several standard stars. Atmospheric extinction and air mass have been taken into account.

\begin{table*}
\begin{center}
\caption{GTC/OSIRIS spectroscopic observations of the six BELLS GALLERY LAEs. \label{tab2}}
\begin{tabular}{c c c c c c c c }
\hline \hline
\smallskip
\smallskip
LAE & PA & Grism &  Rest-frame spectral range & Resolution & Time & Seeing & Depth ($3\sigma$) \\
 & ($^{\circ}$) &  & (\AA) & (km~s$^{-1}$) & (sec) & (arcsec) & ($10^{-17}$~erg~s$^{-1}$~cm$^{-2}$~\AA$^{-1}$) \\
\hline
BG0201+3228 & $-47$ & R2000B; R2500R  & 1034$-$2010 & 230-200 & $2100$; $2800$ & $0.8$ & $0.54-0.19$\\
BG0742+3341 & $63 $ & R2000B  & 1180$-$1695 & 230 & $3600$ & $0.8$ & $0.28$\\ 	
BG0755+3445 & $-30$; $75$ & R2000B; R2500R  & 1086$-$2115 & 230-200 & $7200$; $3600$ & $0.8$ & $0.10-0.16$\\ 
BG0918+5104 & $10.8$ & R2000B; R2500R & 1150$-$2260 & 230-200 & $7200$; $3600$ & $0.7$ & $0.14-0.13$\\ 
BG1429+1202 & $-76.7$ & R2500V; R2500R  & 1150$-$2010 & 160-200 & $3600$; $3600$ & $0.8$ & $0.19-0.17$\\
BG1501+3042 & $29.8$ & R1000B  & 1000$-$2050 & 450 & $3000$ & $1.1$ & $0.20$\\
\hline 
\end{tabular}
\end{center}
\end{table*}

\subsection{Imaging Data}

We carried out imaging observations with the GTC and WHT of five of these lensed LAEs. The lensed LAEs were observed between 2015 April 24 and 2017 May 22 with the Sloan $g$, $r$, and $i$ broad-band filters. The GTC images were obtained with the OSIRIS instrument, with a field of view of $7.8^{\prime} \times 8.5^{\prime}$ and a plate scale of $0.254^{\prime \prime}$ pix$^{-1}$. WHT data were obtained using the Auxiliary-port Camera\footnote{\url{http://www.ing.iac.es/astronomy/instruments/acam/}} \citep[ACAM:][]{benn2008} with a circular field of view of $8^{\prime}$ diameter and a plate scale of $0.254^{\prime \prime}$ pix$^{-1}$. The total integration time ranged between 420 and 900 s for each band, split into several individual exposures (for cosmic rays rejection), under sub-arsec seeing conditions ($\approx 0.7^{\prime \prime} - 0.9^{\prime \prime}$ FWHM).

For all imaging observations, each frame was reduced individually following standard reduction procedures in {\sc Iraf}. These include subtraction of the bias and further correction of the flat-field using, when possible, 
skyflats from low crowded fields. The registration and combination were done using {\sc Scamp} \citep{bertin2006} and {\sc Swarp} \citep{bertin2010}, and the astrometry was improved using Gaia stars from the first data release DR1 \citep{gaia2016}, yielding a typical r.m.s of $\simeq 0.07^{\prime \prime}$. Finally, the images were flux-calibrated against SDSS \citep[Data Release 9 catalog,][]{ahn2012} using stars in the field of view. The GTC images reach $3\sigma$ point source magnitude limits of 25.9, 25.8, and 25.3 mag. in the $g$, $r$, and $i$ filters, respectively. The WHT imaging data reach a depth of 25.5 and 23.3 mag. in the $g$ and $i$ filters, respectively.

In addition, we have used archival $R$ and $I$ wide-field images of BG0918+5104 from MEGACAM on the Canada-France-Hawaii Telescope (CFHT), processed and stacked using the MegaPipe image staking pipeline \citep{gwyn2008}, and downloaded from the Canadian Astronomy Data Centre (CADC
). Total exposure times are 3200 and 2560 s in $R$ and $I$ bands, with an average seeing of $0.85^{\prime \prime}$ and 0.55$^{\prime \prime}$ FWHM, and a $3\sigma$ limiting magnitude of 26.7 and 25.7 (point-source), respectively.

\section{BG1501+3042: confirmation of a new lensed Ly$\alpha$ emitter}\label{sec: lens}

\begin{figure*}
  \centering
   \includegraphics[width=0.75\textwidth]{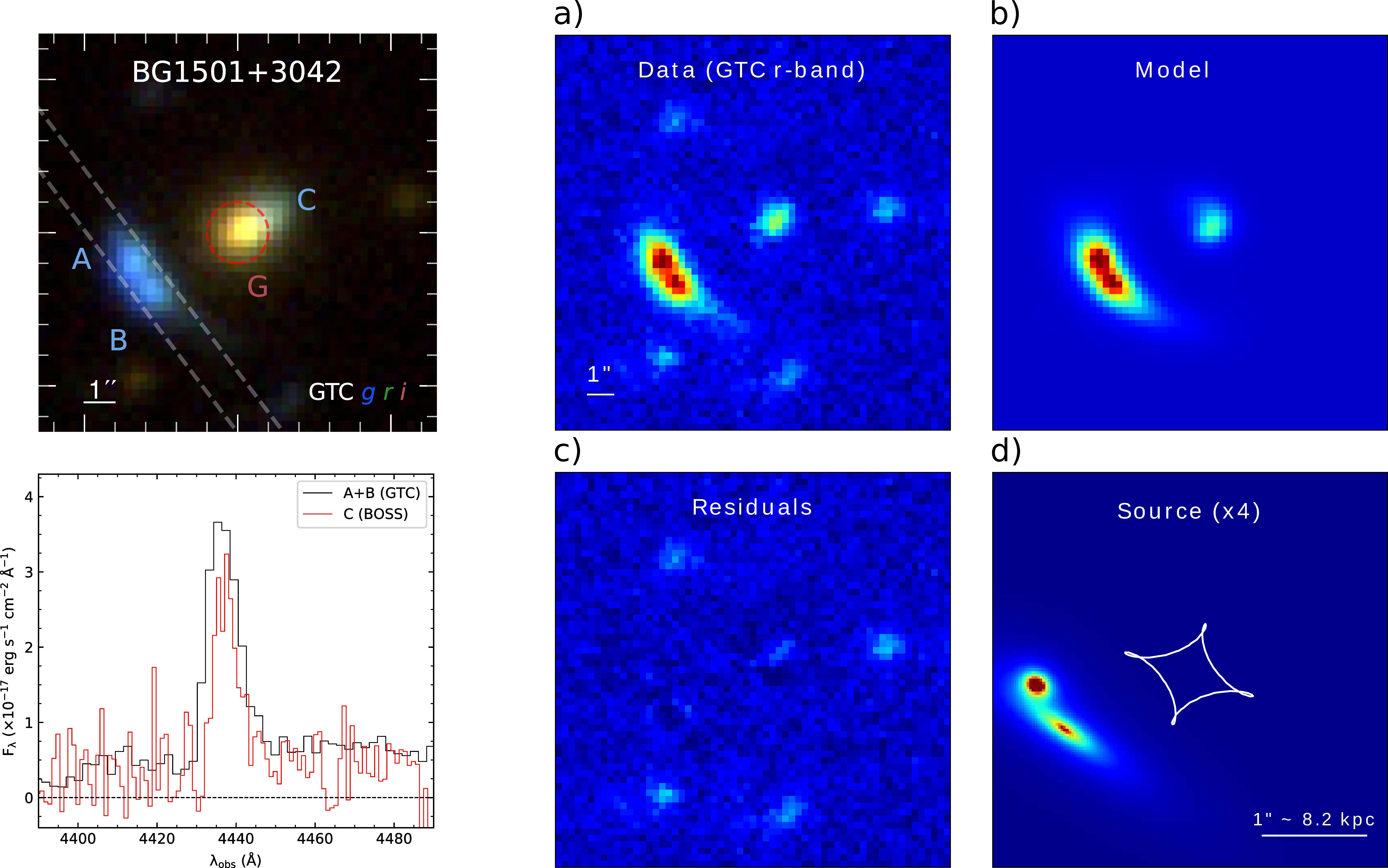}
  \caption{BG1501+3042, a new confirmed lensed LAE. Top left: GTC/OSIRIS rgb color image (using $g-$, $r$-, and $i$-band images). The image is centered on the lensing galaxy (marked as ``G'') and oriented such that north is up and east is to the left; bottom left: spectral region encompassing the Ly$\alpha$ line detected in the lensed images A+B (GTC/OSIRIS, black) and the counter-image C (SDSS/BOSS, red); Right panels: GTC $r$-band with the foreground lens subtracted (a), model with the predicted lensed images (b), final residuals from the best-fit model (c), and position of the two components relative to the caustic (white) in the source plane (d). All panels are centered on the lensing galaxy.}
  \label{fig:bg1501_lens}
\end{figure*}

BG1501+3042 was initially selected by \cite{shu2016a} as a lensed LAE candidate by the detection of an asymmetric emission line at $\simeq 4435$~\AA~(likely Ly$\alpha$ emission at $z=2.649$) in the BOSS spectrum of the $z= 0.638$ LRG SDSS~J150114.60+304230.8. The GTC broad-band images (see Figure \ref{fig:bg1501_lens}) show two bright components at $\simeq 3.7^{\prime \prime}$ SE from the LRG (marked as ``A'' and ``B''), and another one (``C'') partially blended with the LRG (``G''). The GTC spectrum encompassing the bright components ``A'' and ``B'' shows a bright Ly$\alpha$ emission line at $z=2.649$ (see Figure \ref{fig:bg1501_lens}), consistent with the emission line found in the BOSS spectrum of the LRG, that arises from ``C''. Therefore, BG1501+3042 comprises three detected lensed images at the seeing-limited resolution. Such configuration is uncommon in galaxy-scale lensed systems, which typically present two or four lensed images \citep[see, however, other lensed configurations produced by groups or clusters of galaxies, e.g.,][]{oguri2008, dahle2013, shu2018}. 

To understand this system in more detail, we use the 420 s GTC $r$-band image ($\simeq 0.9^{\prime \prime}$ FWHM) to perform the lens modeling for BG1501+3042. We use the same methodology already discussed in previous works \citep[][]{bolton2008b, brownstein2012, shu2015, shu2016c, shu2016b, marques2017}, consisting of a nonlinear optimizer to minimize a $\chi^{2}$ function using the Levenberg-Marquardt algorithm with the {\tt LMFIT} package \citep{lmfit}.
Briefly, the model assumes a mass distribution of the foreground lens parameterized as a singular isothermal ellipsoid (SIE), and an additional external shear is included to model the higher-order effect from the environment. 

The surface brightness distribution of both lens and background source is reconstructed parametrically using elliptical S{\'e}rsic models.
The foreground-light model is combined with the predicted lensed images and convolved with the point-spread function (PSF) that was modeled with stars in the field of view of the GTC/OSIRIS image.

In a first attempt we modelled the system using a single source component. However, the model was not able to reproduce the three detected lensed images of BG1501+3042. To overcome this, we added a second source component to the model. In this case, our best-fit model ($\chi^{2}/\rm d.o.f = 2428/3054$) is able to recover reasonably well the light emission of the three observed lensed images. In this configuration, the lensed images ``A'' and ``B'' are the leading images of the two background sources, and image ``C'' is composed of the blended emission from the two corresponding counter images. The lens model gives a total magnification $\mu = 3.2 \pm 0.5$. Table \ref{all_laes} summarizes the lensing magnification factors derived for the six LAEs studied in this work. 

The projected separation between the centroids of the two components in the source plane is about $0.46^{\prime \prime} \pm 0.06^{\prime \prime}$ which at $z=2.645$ corresponds to $3.8\pm0.6$~kpc for the adopted cosmology, suggestive of a close interacting/merging system. 
Nevertheless, we cannot confirm unambiguously the merging scenario as no velocity offset between the lensed images ``A'' and ``B'' is clearly detected in the available spectrum (with a resolution of $\simeq 450$~km~s$^{-1}$ FWHM).
The sources have effective radius $R_{\rm eff}^{\rm A} =  1.17 \pm 0.47$~kpc, and $R_{\rm eff}^{\rm B} =  3.30 \pm 0.31$~kpc.

\section{Results from Imaging data} \label{sec:image}

\subsection{Photometry}\label{sec:phot}

As seen in Figure \ref{fig:2rgb} the morphology of the LAEs in this sample consists of complex arc-like features with multiple lensed images seen in the observer plane. 

Given the typical $\sim 0.8^{\prime \prime}$ FWHM seeing of the GTC, WHT, and CFHT images, there is light contamination from the foreground lens galaxies. 
In order to perform a clean photometry of the lensed images, we use the two-dimensional fitting program {\sc Galfit} \citep{peng2002, peng2010} to model and substract the light distribution of the foreground lens (and in some cases additionally nearby objects). We use S{\'e}rsic profiles centered at the centroids of each LRG light emission allowing only two pixels freedom ($\sim 0.5^{\prime \prime}$). 
For the four lensed LAEs with $HST$ imaging, we use the models of the foreground lens presented in \cite{shu2016b} as priors in our modeling (i.e. effective radius, minor-to-major axis ratio, and major-axis position angle). Nearby stars were used as PSF models.

\begin{table*}
\begin{center}
\caption{Properties of the six BELLS GALLERY LAEs. \label{all_laes}}
\begin{adjustbox}{max width=1.1\textwidth,center}
\begin{tabular}{l c c c c c c c}
\hline \hline
\smallskip
\smallskip
    & BG0201+3228 & BG0742+3341 & BG0755+3445 & BG0918+5104 & BG1429+1202 & BG1501+3042 & Notes\\
\hline
R.A.  & 02:01:21.39 & 07:42:49.68 & 07:55:23.52 & 09:18:59.21 & 14:29:54.80 & 15:01:14.61 & (1)\\
Dec.  & +32:28:29.7 & +33:41:49.0 & +34:45:39.6 & +51:04:52.6 & +12:02:35.6 & +30:42:30.8 & (1)\\
$z_{\rm lens}$  & $0.3957 \pm 0.0001$ & $0.4936\pm0.0001$ & $0.7223\pm0.0002$ & $0.5811\pm0.0002$ & $0.5531\pm0.0002$ & $0.6384\pm0.0002$ & (2)\\
$z_{\rm LAE}$  & $2.8174\pm0.0007$ & $2.3625\pm0.0009$ & $2.6345\pm0.0005$ & $2.4000\pm0.0003$ & $2.8244\pm0.0006$ & $2.645\pm0.001$ & (2)\\
$g$  & $20.94 \pm 0.09$ & $-$ & $21.68 \pm 0.10$ & $21.69 \pm 0.09$ & $20.40 \pm 0.08$ & $21.32 \pm 0.10$ & (3)\\
$r$   & $20.62 \pm 0.08$ & $-$ & $21.16 \pm 0.08$ & $21.01 \pm 0.09$ & $20.14 \pm 0.05$ & $21.00 \pm 0.05$ &  (3)\\
$i$  & $20.47 \pm 0.20$ $^{\rm a)}$ & $-$ & $21.04 \pm 0.06$ & $20.91 \pm 0.09$ & $20.08 \pm 0.05$ & $20.98 \pm 0.07$ & (3)\\
F606W  & $20.56 \pm 0.05$ & $22.18 \pm 0.05$ & $21.11 \pm 0.05$ & $20.87 \pm 0.05$ & $-$ & $-$ & (3)\\
$\mu$  & $15\pm3$ & $16\pm3$ & $14\pm2$ & $18\pm3$ & $8.8\pm1.4$ & $3.2\pm0.5$ & (4)\\
$\beta_{\rm UV}$  & $-1.50 \pm 0.98$ & $-$ & $-1.59 \pm 0.45$ & $-1.58 \pm 0.59$ & $-1.76 \pm 0.32$ & $-1.96 \pm 0.39$ & (5)\\
$A_{\rm UV}$  & $1.71\pm1.71$ & $-$ & $1.55\pm 0.82$ & $1.56\pm 1.08$ & $1.24\pm 0.58$ & $0.87\pm 0.72$ & (5)\\
{\it E(B-V)}  & $0.20\pm 0.20$ & $-$ & $0.18\pm 0.10$ & $0.19\pm 0.13$ & $0.14\pm 0.07$ & $0.10\pm 0.08$ & (5)\\
$F^{\rm GTC}_{\rm Ly\alpha}$   & $7.5\pm0.3$ & $7.1\pm0.2$ & $5.2\pm0.2$ & $3.0\pm0.2$ & $11.5\pm0.2$ & $4.1\pm0.2$ & (6)\\
$EW_{0}$~(Ly$\alpha$)  & $23\pm6$ & $45\pm6$ & $28\pm5$ & $16\pm5$ & $49\pm7$ & $16\pm5$ & (6)\\
FWHM~(Ly$\alpha$) & $395\pm70$ & $230\pm50$ & $330\pm60$ & $260\pm60$ & $290\pm40$ & $340 \pm 150$ & (6) \\
$\Delta v$~(Ly$\alpha$)  & $230\pm80$ & $80\pm70$ & $80\pm70$ & $210\pm80$ & $30\pm 60$ & $290\pm 120$ & (7)\\
$\Delta v$~(ISM)  & $-100\pm70$ & $-20\pm80$ & $-150\pm80$ & $-10\pm70$ & $-230\pm60$ & $-175\pm100$ & (7)\\
$C_{f}$~(low-ISM) & $0.67\pm0.15$ & $>0.51$ & $0.66\pm0.17$ & $0.53\pm0.11$ & $0.44\pm0.08$ & $0.35\pm0.17$ & (8) \\
$C_{f}$~(H~{\sc i}) & $0.96\pm0.17$ & $>0.93$ & $0.95\pm0.17$ & $0.83\pm0.14$ & $0.68\pm0.13$ & $0.72\pm0.13$ & (8) \\
$M_{\rm UV}$  & $-22.06 \pm 0.24$ & $-19.95 \pm 0.23$ & $-21.43 \pm 0.23$ & $-21.05 \pm 0.22$ & $-23.04 \pm 0.21$ & $-23.12 \pm 0.21$ & (9)\\
SFR (UV)  & $110_{-80}^{+470}$ & $\gtrapprox 3$ & $55_{-30}^{+60}$ & $40_{-25}^{+65}$ & $175_{-70}^{+125}$ & $135_{-55}^{+115}$ & (9)\\
$L_{\rm Ly \alpha}$  & $4.43\pm0.74$ & $2.02\pm0.32$ & $2.48\pm0.31$ & $1.36\pm0.19$ & $30.67\pm4.21$ & $14.24\pm1.93$ & (10)\\
SFR (Ly$\alpha$)  & $15.7_{-13.4}^{+103.3}$ & $\gtrapprox 1.0$ & $7.3_{-4.5}^{+11.3}$ & $4.1_{-2.9}^{+9.7}$ & $63.2_{-30.7}^{+59.3}$ & $17.4_{-9.7}^{+21.8}$ & (10)\\
$f_{\rm esc}$~(Ly$\alpha$)  & $0.14\pm0.02$ & $0.37\pm0.02$ &   $0.14\pm 0.01$ & $0.11\pm0.01$ &  $0.36\pm 0.01$ &$0.14\pm0.01$ & (11)\\
\hline 
\end{tabular}
\end{adjustbox}
\end{center}
\textbf{Notes. ---} (1) positions of the lens galaxy (J2000); (2) systemic redshifts of the lens \citep{shu2016a} and LAEs (Section \ref{sys}); (3) integrated magnitudes of the LAEs in each band (AB); (4) lensing magnification factors derived in \cite{shu2016b} for the first four lensed LAEs, \cite{marques2017} for BG1429+1202, and in Section \ref{sec: lens} for BG1501+3042. (5) stellar UV continuum slope (equation \ref{eq_beta_slope}), UV dust extinction (AB mag, equation \ref{eq_dust_ext}), and color excess of the stellar continuum assuming the \cite{reddy2015} attenuation curve; (6) observed flux (units of $10^{-16}$~erg~s$^{-1}$~cm$^{-2}$), rest-frame equivalent width (\AA), and full width half maximum (km~s$^{-1}$) of the Ly$\alpha$ line; (7) centroid velocity of Ly$\alpha$ and ISM absorption lines (average) relative to the systemic redshift (km~s$^{-1}$); (8) low-ISM and H~{\sc i} covering fractions using the depth of the average absorption line profile of ISM lines (Si~{\sc ii}~1260\AA, C~{\sc ii}~1334\AA, and Si~{\sc ii}~1526\AA) and using the relation of \cite{gazagnes2018}, respectively; (9)  intrinsic UV absolute magnitudes (AB) and star formation rates ($M_{\odot}$~yr$^{-1}$; equation \ref{eq_sfruv}); (10) intrinsic Ly$\alpha$ luminosity (units of $10^{42}$~erg~s$^{-1}$) and star formation rates ($M_{\odot}$~yr$^{-1}$; equation \ref{eq_lya}); and (11) Ly$\alpha$ escape fraction. \\

$^{\rm a)}$ The ACAM/WHT $i$ band data of BG0201+3228 are affected by a high background level. 
\end{table*}

We further use {\sc SExtractor} \citep{bertin1996} to measure the magnitudes in each band. 
To do so, we use aperture photometry using apertures with diameters of 2.5 times the FWHM of the PSF. Aperture photometry was performed for compact lensed morphologies, such as the lensed images of BG0755+3445, BG1429+1202 and the faint lensed counter-images of BG0201+3228 or BG1501+3042. For more complex morphologies, such as bright extended arcs seen in BG0201+3228, BG0918+5104 and BG1501+3042, we derived total magnitudes using the corresponding {\sc SExtractor}'s \texttt{MAG AUTO} output.
Table \ref{all_laes} summarizes the photometric measurements for each lensed LAE. 

\subsection{UV Luminosity}
 
Knowing the total magnification of these galaxies, we determine the intrinsic absolute UV magnitude ($M_{\rm UV}$ in units of AB magnitudes) in the rest-frame UV, typically defined at $\approx 1500 - 1700$ \AA: 

\begin{equation}
    M_{UV} = m - 5  \log_{10} (D_{ L}) + 5 + 2.5 \log_{10} (1 + { z}) + 2.5 \log_{10} (\mu),
\end{equation}

\noindent
where $m$ is the apparent AB magnitude in the closest observed band to the rest-frame 1600~\AA{ }\citep[corrected for the Galactic dust extinction using the values from][]{schlafly2011}, $D_{L}$ is the luminosity distance in pc at a given redshift ($z$), and $\mu$ is the total magnification factor. 

We use the $HST$ F606W band ($\lambda_{\rm eff} = 5778$ \AA) to trace the rest-frame $1600$~\AA{ }of BG0742+3341 ($z = 2.363$) and BG0918+5104 ($z = 2.400$), whereas for the highest-redshift LAEs we use the ground-based $r$-band apparent magnitudes ($\lambda_{\rm eff} = 6231$ \AA). 
These LAEs span a wide range of $M_{\rm UV}$, from $-19.9$ to $-23.1$~AB (see Table \ref{all_laes}).

\subsection{Beta Slope, Dust Attenuation and Star Formation Rate}\label{sfr_uv}

The stellar UV continuum slope $\beta_{\rm UV}$ is a useful indicator of the dust content in star-forming 
galaxies, in the sense that larger quantities of dust produce redder UV SEDs. Assuming a simple power law $f_{\lambda} \propto \lambda^{\beta_{\rm UV}}$, the slope can be estimated as:

\begin{equation}\label{eq_beta_slope}
\beta_{\rm UV} = - \frac{m_{1} - m_{2}}{2.5 \log_{10} (\lambda_{m_{1}} / \lambda_{m_{2}})} - 2.  
\end{equation}

\noindent
We use the $r$ ($m_{1}$, $\lambda_{1} = 6231$ \AA) and $i$ ($m_{2}$, $\lambda_{2} = 7625$ \AA) bands, which at $z = 2.4 - 2.8$ correspond to a rest-frame spectral coverage of about $1600 - 2200$ \AA. 
Following \cite{reddy2015}, $\beta_{\rm UV}$ is correlated with the dust extinction at $\lambda = 1600$~\AA, $A_{\rm UV}$, in the form:

\begin{equation}\label{eq_dust_ext}
A_{\rm UV} = 1.84 (\beta_{\rm UV} - \beta_{\rm UV,0}) = k_{\rm UV} \times E(B-V),  
\end{equation}

\noindent
where $\beta_{\rm UV,0} = -2.44$ and represents the intrinsic UV spectral slope in the absence of dust, $k_{\rm UV}$ is the attenuation coefficient at $1600$~\AA{ }($\simeq 8.34$; \citealt{reddy2015}), and {\it E(B-V)} is the color excess of the stellar continuum. Although not relevant here (see Section \ref{sec:lya}), we consider similar {\it E(B-V)} for stellar and nebular components \citep[e.g.,][]{koyama2015, faisst2019}.

Finally, using the Kennicutt's conversion \citep{kennicutt1998}, the rest-frame UV luminosity (already 
corrected for the lensing magnification) can be translated into an intrinsic UV star formation rate (in units of $M_{\odot}$ yr$^{-1}$): 

\begin{equation}\label{eq_sfruv}
\rm SFR (UV)  = 
1.48 \times 10^{-28} \times L_{\rm UV} \times 10^{0.4 A_{\rm UV}}.  
\end{equation}

Note that, however, this expression assumes the standard \cite{salpeter1955} stellar initial mass function (IMF). In order to  correct for the lower proportion of low-mass stars \citep[e.g.,][IMF]{chabrier2003}, we divide the SFR by $1.8$. 

Measurements of $\beta_{\rm UV}$, $A_{\rm UV}$, {\it E(B-V)}, and lensing- and dust-corrected SFRs~(UV) for all LAEs are listed in Table \ref{all_laes}. 
For comparison, adopting the \cite{calzetti2000} extinction law we would find values of $A_{\rm UV}$ slightly larger (3\% to 15\%, depending on each LAE) than the ones measured using Equation \ref{eq_dust_ext}, yielding also to larger SFRs (3\% to 25\%). On the other hand, some authors argue that young and low-mass galaxies are better modeled with an extinction curve close to the Small Magellanic Cloud \citep[SMC; e.g.,][]{bouwens2016, reddy2018, alvarez2019}. Adopting the SMC extinction curve, we would find SFRs lower by a factor of $\approx 2$.
 
\section{Rest-frame UV Spectra} \label{sec:spec}

\begin{figure*}
  \centering
   \includegraphics[width=0.99\textwidth]{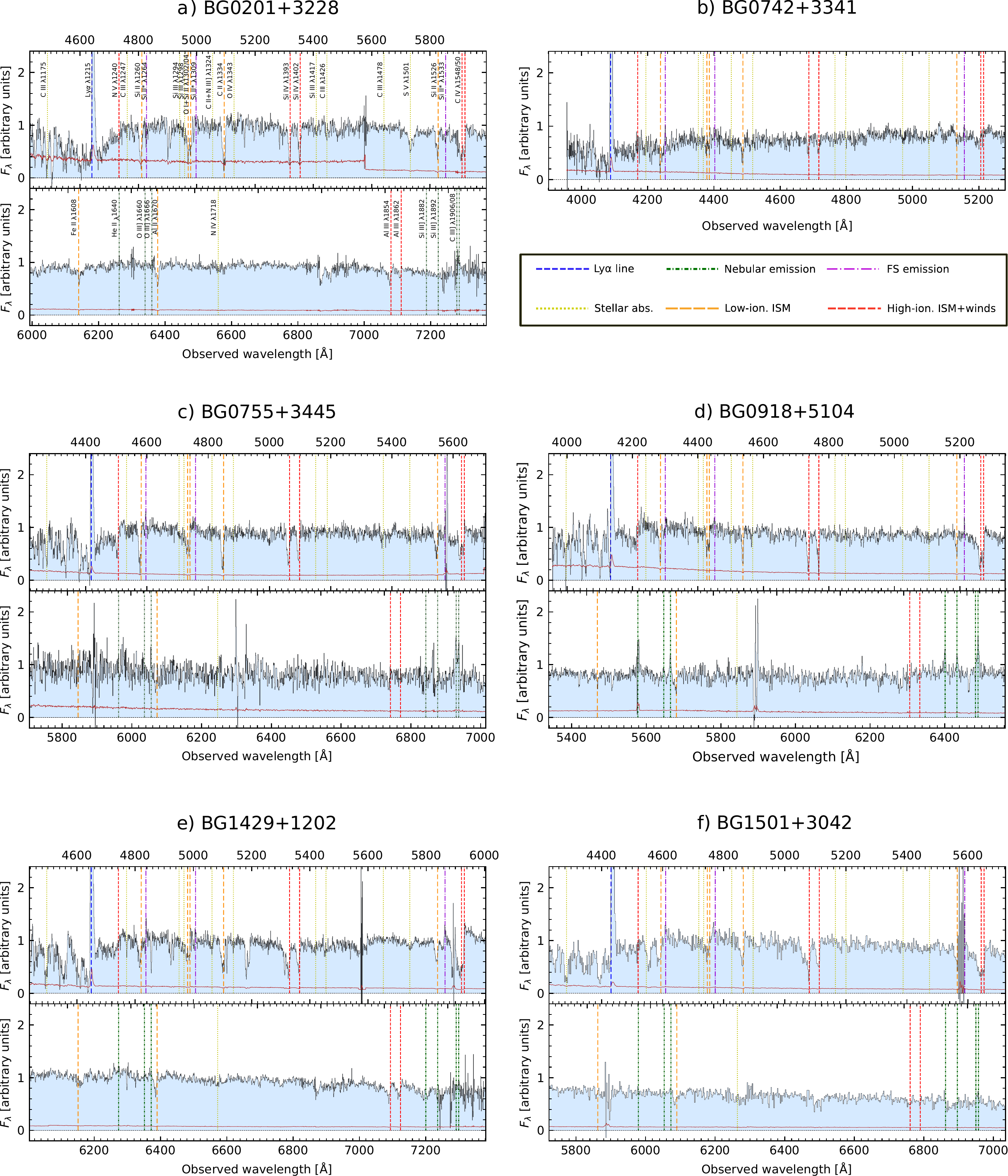}
  \caption{Portions of the rest-frame UV spectra of the 6 LAEs observed with OSIRIS. From panels a) to f) we show respectively the spectra of BG0201+3228, BG0742+3341, BG0755+3445, BG0918+5104, BG1429+1202, and BG1501+3042. Ticks mark the position of the Ly$\alpha$ line (blue) as well as other main features associated with nebular emission (green), fine-structure emission (purple), stellar photospheric absorption (yellow), low-ionization ISM absorption (orange), and high-ionization ISM absorption and stellar winds (P-Cygni profiles, red). The $1 \sigma$ uncertainty (per spectral element) is plotted in red.}
  \label{fig:4}
\end{figure*}

In this Section we analyze the rest-frame UV spectra of these lensed LAE systems. 
All spectra, shown in Figure \ref{fig:4}, have relatively high S/N per pixel in the continuum, despite the short exposure time, ranging from $\sim 7$ for the faintest LAE (BG0742+3341) to $\sim 20$ for the brightest ones. 
These spectra show prominent Ly$\alpha$ emission spatially coincident with the lensed images of the background LAEs. 
The achieved S/N is also sufficient to detect a wealth of other absorption/emission features associated to the photospheres and winds of massive stars, neutral and slightly ionized interstellar medium gas (ISM), and ionized gas in H~{\sc ii} regions.
The spectrum of BG0742+3341 exhibits a redder UV slope than in the other LAEs, suggesting a significant foreground light contamination from the lens galaxy (this LRG is bright and is relatively close to the arc, $\lesssim 1^{\prime \prime}$). 
For the remaining LAEs, in particular for BG1429+1202 and BG1501+3042, we do not expect a large light contamination in their spectra ($<10$\%), since the bright lensed images covered by the GTC long-slits are faraway from the lens galaxies ($\simeq 1.7^{\prime \prime} - 3^{\prime \prime}$).

\subsection{Galaxy Systemic Redshifts}\label{sys}

\begin{figure}
  \centering
  \includegraphics[width=0.46\textwidth]{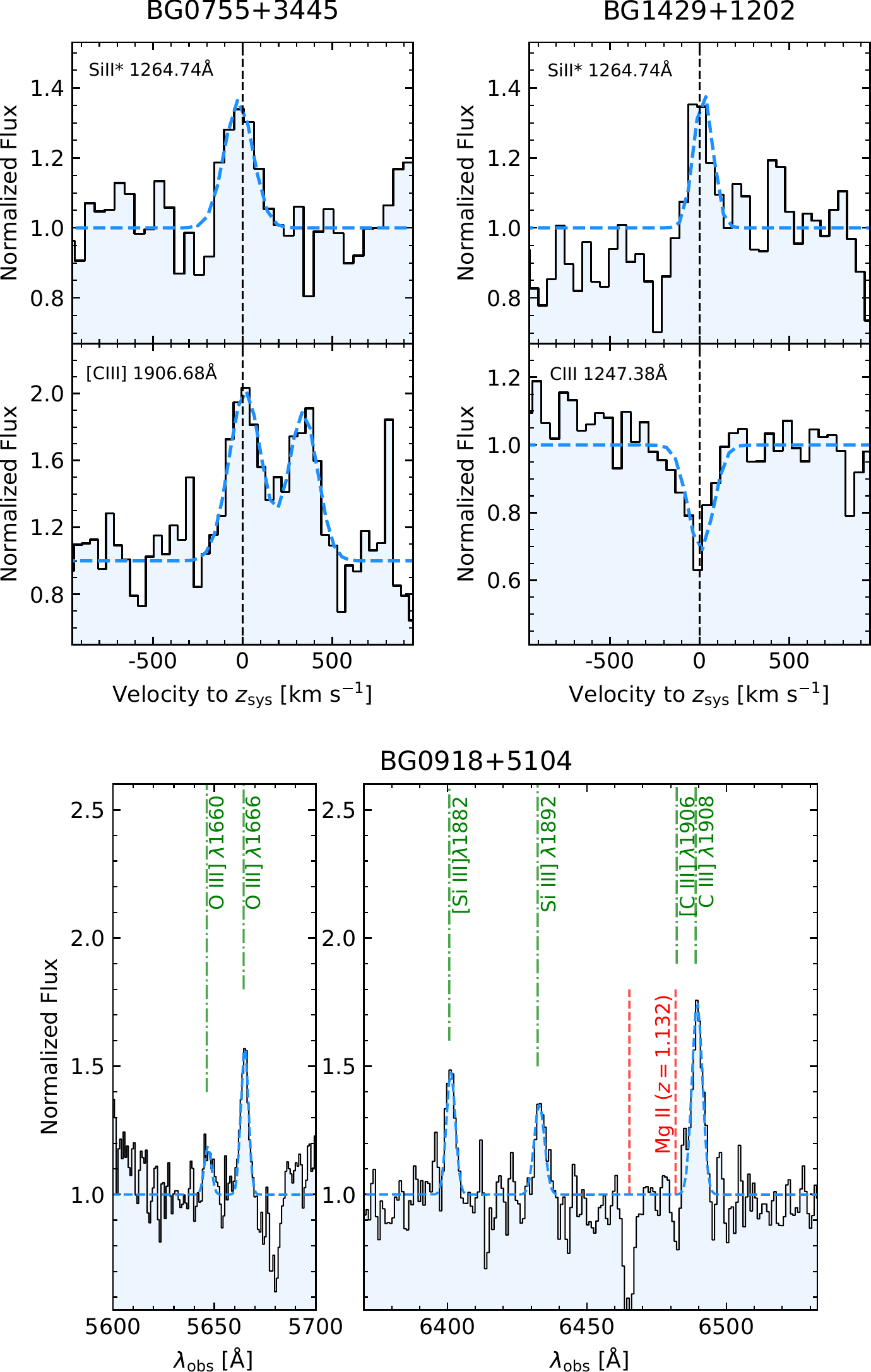}
  \caption{Determination of the systemic redshifts. \textit{Upper}: comparison between the spectral profiles on rest velocity of fine-structure Si~{\sc ii*}~1264\AA{ }emission lines (upper panels) in BG0755+3445 and BG1429+1202 (left and right, respectively) and those of  H~{\sc ii} emission ([C~{\sc iii}]~1906\AA{ }, bottom left) and stellar absorption (C~{\sc iii}~1247\AA, bottom right). Fine-structure emission lines agree with the nebular and stellar redshifts within a rms of $\lesssim 50$~km~s$^{-1}$. \textit{Bottom:} nebular emission lines in BG0918+5104 for which the systemic redshift of $z = 2.4000 \pm 0.0003$ was measured. [C~{\sc iii}]~1906\AA{ }nebular emission is highly affected by the absorption of the Mg~{\sc ii} $\lambda$2796,2803 doublet from an intervening absorption system at $z = 1.312 \pm 0.001$ (see Section \ref{low-z}).}
  \label{fig:fs}
\end{figure}

Systemic redshifts are measured using stellar absorption lines, generated in the photospheres of stars, and nebular emission from H~{\sc ii} regions. 
In the absence of a clear detection of nebular or stellar features we also use the Si~{\sc ii*} fine-structure emission lines in deriving the systemic redshift.
These lines (Si~{\sc ii*}~1197\AA, Si~{\sc ii*}~1264\AA, and Si~{\sc ii*}~1533\AA) are produced by resonant scattering and re-emission of ground-state interstellar absorption features, and 
as already shown in \cite{france2010}, \cite{steidel2016} and \cite{jones2018}, they are useful for the determination of the galaxy rest velocity when observed with significant high spectral resolution, avoiding the contamination from the neighboring red component of resonance absorption features (Si~{\sc ii}~1193\AA, Si~{\sc ii}~1260\AA, and Si~{\sc ii}~1526\AA, respectively). This is demonstrated in the upper panels of Figure \ref{fig:fs}, where we compare the spectral profiles of Si~{\sc ii*}~1264\AA{ }in two LAEs (BG0755+3445 and BG1429+1202, left and right, respectively) to those of nebular emission ([C~{\sc iii}]~1906\AA) and stellar photospheric absorption (C~{\sc iii}~1247\AA) on the systemic velocity of each galaxy. Si~{\sc ii*} emission lines agree with the nebular and stellar redshifts within a rms of $\lesssim 50$~km~s$^{-1}$.

We now list the most prominent lines used for the measurements of the systemic redshifts of each LAE. 

BG0201+ 3228 shows several faint stellar photospheric absorption lines including C~{\sc iii}~1247\AA, Si~{\sc iii}~1294,1298\AA, and C~{\sc iii}~1426\AA. We use the mean of them to derive the rest-velocity of BG0201+3228 as $z_{\rm sys} = 2.8174 \pm 0.0007$. Nebular O~{\sc iii}]~1666\AA{ }and C~{\sc iii}]~1906,1908\AA{ }emission lines are only barely detected ($\sim 2 \sigma$; they fall in regions of relatively strong night sky lines), making their measurements difficult.

For BG0742+3341, due to the limited wavelength range of its spectrum, from $1180$ to $1700$ \AA~\space in the rest-frame (it was not observed with the R2500R grism), and the relatively modest S/N continuum spectrum (BG0742+3341 is the faintest LAE in our sample), we do not detect clearly any photospheric stellar absorption line or nebular emission.  
Nevertheless, we clearly detect three fine-structure emission lines, Si~{\sc ii*}~1197\AA, Si~{\sc ii*}~1264\AA, and Si~{\sc ii*}~1533\AA, at a redshift $z = 2.3625 \pm 0.0009$.  
We thus assume the redshift of the fine-structure emission lines as the systemic redshift of BG0742+3341.

The combined spectrum of BG0755+3445 (covering the four lensed images, A to D in Figure \ref{fig:2rgb}) shows a prominent and resolved C~{\sc iii]}~1906,1908\AA{ }nebular emission at $z_{\rm neb} = 2.6345 \pm 0.0005$ (see Figure \ref{fig:fs}), as well as faint stellar photospheric absorption lines, such as C~{\sc iii}~1247\AA{ }and Si~{\sc iii}~1298\AA, at $z_{\rm stars} = 2.6340 \pm 0.0007$. We use both nebular emission and stellar absorption lines to measure the systemic redshift $z_{\rm sys} = 2.6342 \pm 0.0008$.

For BG0918+5104, several nebular emission lines are detected with high significance ($\sigma \gtrapprox 5$), including the O~{\sc iii}]~1660,1666\AA, and Si~{\sc iii}]~1882,1892\AA{ }doublets, for which we derive the systemic redshift of $z_{\rm sys} = 2.4000 \pm 0.0003$ (see Figure \ref{fig:fs}). 
The nebular C~{\sc iii}]~1908\AA{ }emission is also detected with high significance, but not the corresponding [C~{\sc iii}]~1906\AA{ }line, whose profile is highly affected by the absorption of the Mg~{\sc ii}~2796,2803\AA{ }doublet from an intervening absorption system at $z = 1.312 \pm 0.001$ (see Appendix \ref{low-z} for more details). 
In addition, the fine-structure emission line Si~{\sc ii*}~1533\AA{ }is also detected at $z_{\rm FS} = 2.4004 \pm 0.0006$, consistent with the systemic redshift from H~{\sc ii} regions.

The new spectroscopic observations of BG1429+2102 allow us to identify several photospheric absorption features, C~{\sc iii}~1247\AA, Si~{\sc iii}~1294,1298\AA, the close blend of C~{\sc ii} and N~{\sc iii}~1324\AA, S~{\sc v}~1501\AA, and N~{\sc iv}~1718\AA, and redefine the systemic redshift of BG1429+1202 as $z_{\rm sys} = 2.8244 \pm 0.0006$, similar to the one measured by \cite{rigby2017b} ($z_{\rm sys} = 2.8241 \pm 0.0006$).
Emission in nebular lines of O~{\sc iii]}~1666\AA{ }and C~{\sc iii]}~1906,1908\AA{ }is also detected, but with low significance. 
Fine-structure emission lines Si~{\sc ii*}~1264\AA, Si~{\sc ii*}~1309\AA, and Si~{\sc ii*}~1533\AA{ }are also clearly detected and their peak lie at $z_{\rm FS} = 2.8243 \pm 0.0008$, consistent with the systemic redshift derived using photospheric absorption lines (see top right panel of Figure \ref{fig:fs}).

Finally, despite the relatively modest S/N continuum and the low spectral resolution provided by the R1000B grism ($\sim 450$ km s$^{-1}$), the spectrum of BG1501+3042 shows both stellar absorption lines (Si~{\sc iii}~1294\AA, C~{\sc ii} and Si~{\sc iii}~1296\AA, and Si~{\sc iii}~1298\AA), and nebular emission lines in O~{\sc iii]}~1666\AA{ }and C~{\sc iii}]~1906,1908\AA{ }for which we measure the systemic redshift $z_{\rm sys} = 2.645 \pm 0.001$.

\subsection{ISM Absorption Lines}\label{ism}

The strongest absorption features seen in the spectra of these LAEs in Figure \ref{fig:4} are associated with the interstellar medium gas (ISM) and are produced by the resonance transition of several ionic species (e.g., H, C, O, Al, Si, Fe, among others). 
We detect several ISM absorption lines in a variety of ionization stages, from neutral or partially neutral (e.g., O~{\sc i}, Si~{\sc ii}, C~{\sc ii}), to highly ionized species (e.g., Al~{\sc iii}, Si~{\sc iv}, or C~{\sc iv}), which predominantly trace gas at higher temperatures. 

Since ISM absorption lines are seen against the continuum provided by the integrated light of O and B stars in a galaxy, they offer a unique probe of the kinematics of the gas. 
It is well accepted that large-scale galactic outflows are ubiquitous in nearly all star-forming galaxies due the kinetic energy deposited by stellar and supernova-driven winds. Such signatures have been evaluated from blueshifted ISM absorption lines with respect to the systemic redshift \citep[e.g.,][]{shapley2003, weiner2009, steidel2010}, i.e., the gas on the near side of the galaxy is moving towards the observer.

\begin{table*}
\begin{center}
\caption{ISM Absorption Line Features. \label{table_abs_line} }
\begin{adjustbox}{totalheight=\textheight-7\baselineskip}
\begin{tabular}{c c c c c c c}
\hline \hline
\smallskip
\smallskip
LAE$^{\rm (1)}$ & Ion$^{\rm (2)}$ & $\lambda_{\rm rest}^{\rm (3)}$ & $\Delta v^{\rm (4)}$  & FWHM$^{\rm (5)}$ & $EW_{0}^{\rm (6)}$ & $I/I_0^{\rm (7)}$ \\ 
 &  &  (\AA) & (km~s$^{-1}$) & (km~s$^{-1}$) &   (\AA) &    \\
\hline
BG0201+3228	& Si {\sc ii} & 1260.42 & $-155\pm60$ & $480\pm65$ & $2.52\pm0.31$ & $0.17\pm0.09$  \\ 
            & O {\sc i} & 1302.16 & $-21\pm80$ & $510\pm80$ & $2.31\pm0.30$ & $0.27\pm0.08$ \\ 
            & Si {\sc ii} & 1304.37 & $-85\pm80$ & $440\pm80$ & $1.64\pm0.26$ & $0.48\pm0.06$ \\ 
            & C {\sc ii} & 1334.53 & $-20\pm60$ & $540\pm65$ & $3.07\pm0.32$ & $0.13\pm0.10$ \\
	    	& Si {\sc iv} & 1393.76 & $-90\pm60$ & $350\pm70$ & $1.65\pm0.23$ & $0.23\pm0.09$ \\
	    	& Si {\sc iv} & 1402.77 & $-70\pm60$ & $375\pm70$ & $1.52\pm0.24$ & $0.24\pm0.09$ \\
		    & Si {\sc ii} & 1526.70& $-126\pm60$ & $415\pm75$ & $1.61\pm0.21$ & $0.34\pm0.07$ \\	    	
	    	& C {\sc iv} & 1548.20 & $-240\pm9$ & $560\pm90$ & $2.27\pm0.28$ & $0.35\pm0.07$ \\
	    	& C {\sc iv} & 1550.78 & $-80\pm90$ & $280\pm90$ & $1.21\pm0.18$ & $0.47\pm0.07$ \\
	    	& Fe {\sc ii} & 1608.45 & $-300\pm90$ & $<200$ & $0.50\pm0.08$ & $0.65\pm0.05$\\
	     	& Al {\sc ii} & 1670.78 & $-40\pm60$ & $285\pm80$ & $1.11\pm0.15$ & $0.57\pm0.05$ \\ 
            & Al {\sc iii}& 1854.72 & $-130\pm75$ & $285\pm80$ & $0.62\pm0.12$ & $0.58\pm0.05$ \\
            & Al {\sc iii}& 1862.80 & $-120\pm75$ & $<200$ & $0.27\pm0.05$ & $0.76\pm0.05$ \\ 
\hline
BG0742+3341	& Si {\sc ii} & 1260.42 & $-35\pm80$ & $<230$ & $0.62\pm0.14$ & $0.21\pm0.19$  \\ 
            & O {\sc i} & 1302.16 & $-100\pm90$ & $<230$ & $0.41\pm0.10$ & $0.66\pm0.18$ \\ 
            & Si {\sc ii} & 1304.37 & $-85\pm90$ & $<230$ & $0.30\pm0.08$ & $0.71\pm0.18$ \\ 
            & C {\sc ii} & 1334.53 & $15\pm70$ & $<230$ & $0.60\pm0.13$ & $0.42\pm0.15$ \\
	    	& Si {\sc iv} & 1393.76 & $-160\pm70$ & $190\pm70$ & $0.58\pm0.13$ & $0.64\pm0.13$ \\
	    	& Si {\sc iv} & 1402.77 & $-50\pm70$ & $280\pm80$ & $0.62\pm0.15$ & $0.68\pm0.13$ \\
		    & Si {\sc ii} & 1526.70& $-70\pm80$ & $<230$ & $0.17\pm0.06$ & $0.77\pm0.21$ \\	    	
	    	& C {\sc iv} & 1548.20 & $-320\pm100$ & $300\pm100$ & $0.58\pm0.16$ & $0.71\pm0.18$ \\
	    	& C {\sc iv} & 1550.78 & $-200\pm100$ & $<230$ & $0.14\pm0.05$ & $0.67\pm0.18$ \\ 
\hline
BG0755+3445	& Si {\sc ii} & 1260.42 & $-180\pm70$ & $415\pm80$ & $1.62\pm0.19$ & $0.18\pm0.09$  \\ 
            & O {\sc i} & 1302.16 & $-65\pm60$ & $365\pm70$ & $0.91\pm0.09$ & $0.55\pm0.05$ \\ 
            & Si {\sc ii} & 1304.37 & $-140\pm65$ & $<230$ & $0.65\pm0.07$ & $0.49\pm0.06$ \\ 
            & C {\sc ii} & 1334.53 & $-200\pm60$ & $430\pm70$ & $1.78\pm0.19$ & $0.24\pm0.08$ \\
	    	& Si {\sc iv} & 1393.76 & $-220\pm65$ & $480\pm80$ & $1.56\pm0.16$ & $0.33\pm0.07$ \\
	    	& Si {\sc iv} & 1402.77 & $-225\pm65$ & $360\pm70$ & $1.16\pm0.13$ & $0.41\pm0.07$ \\
		    & Si {\sc ii} & 1526.70& $-120\pm65$ & $480\pm70$ & $1.25\pm0.14$ & $0.48\pm0.06$ \\	    	
	    	& C {\sc iv} & 1548.20 & $-320\pm90$ & $465\pm90$ & $1.45\pm0.16$ & $0.45\pm0.06$ \\
	    	& C {\sc iv} & 1550.78 & $-290\pm90$ & $280\pm90$ & $0.62\pm0.07$ & $0.55\pm0.05$ \\ 
\hline
BG0918+5104	& Si {\sc ii} & 1260.42 & $-15\pm60$ & $385\pm80$ & $1.27\pm0.21$ & $0.38\pm0.11$  \\ 
            & O {\sc i} & 1302.16 & $-10\pm65$ & $200\pm85$ & $0.70\pm0.12$ & $0.49\pm0.10$ \\ 
            & Si {\sc ii} & 1304.37 & $-80\pm65$ & $290\pm80$ & $0.62\pm0.10$ & $0.63\pm0.06$ \\ 
            & C {\sc ii} & 1334.53 & $-3\pm60$ & $240\pm60$ & $1.09\pm0.18$ & $0.33\pm0.12$ \\
	    	& Si {\sc iv} & 1393.76 & $-120\pm60$ & $350\pm70$ & $1.44\pm0.25$ & $0.26\pm0.12$ \\
	    	& Si {\sc iv} & 1402.77 & $-85\pm65$ & $320\pm70$ & $1.29\pm0.22$ & $0.23\pm0.13$ \\
		    & Si {\sc ii} & 1526.70& $15\pm65$ & $250\pm70$ & $0.75\pm0.13$ & $0.55\pm0.07$ \\	    	
	    	& C {\sc iv} & 1548.20 & $-170\pm100$ & $550\pm150$ & $2.04\pm0.4$ & $0.28\pm0.12$ \\
	    	& C {\sc iv} & 1550.78 & $-160\pm80$ & $<200$ & $0.74\pm0.15$ & $0.37\pm0.11$ \\ 
	    	& Al {\sc ii} & 1670.78 & $-6\pm60$ & $<200$ & $0.73\pm0.15$ & $0.68 \pm0.11$ \\ 
\hline
BG1429+1202	& Si {\sc ii} & 1260.42 & $-250\pm55$ & $550\pm55$ & $1.28\pm0.10$ & $0.59\pm0.07$  \\ 
            & O {\sc i} & 1302.16 & $-160\pm60$ & $435\pm60$ & $0.81\pm0.09$ & $63\pm0.04$ \\ 
            & Si {\sc ii} & 1304.37 & $-160\pm60$ & $255\pm60$ & $0.54\pm0.06$ & $0.61\pm0.04$ \\ 
            & C {\sc ii} & 1334.53 & $-170\pm55^{\rm a)}$ & $710\pm80^{\rm a)}$ & $1.47\pm0.13^{\rm a)}$ & $0.40\pm0.05^{\rm a)}$ \\
	    	& Si {\sc iv} & 1393.76 & $-250\pm50^{\rm a)}$ & $750\pm80^{\rm a)}$ & $2.01\pm0.21^{\rm a)}$ & $0.34\pm0.07^{\rm a)}$ \\
	    	& Si {\sc iv} & 1402.77 & $-240\pm50$ & $495\pm50$ & $1.13\pm0.10$ & $0.38\pm0.06$ \\
		    & Si {\sc ii} & 1526.70& $-205\pm50$ & $460\pm60$ & $1.07\pm0.09$ & $0.61\pm0.05$ \\	    	
	    	& C {\sc iv} & 1548.20 & $-590\pm90^{\rm b)}$ & $1100\pm150^{\rm b)}$ & $4.28\pm0.41^{\rm b)}$ & $0.25\pm0.07^{\rm b)}$ \\
	    	& C {\sc iv} & 1550.78 & $-590\pm90^{\rm b)}$ & $1100\pm150^{\rm b)}$ & $4.28\pm0.41^{\rm b)}$ & $0.25\pm0.07^{\rm b)}$ \\ 
	    	& Al {\sc ii} & 1670.78 & $-195\pm50$ & $320\pm50$ & $0.68\pm0.07$ & $0.71\pm0.05$ \\ 
            & Al~{\sc iii}& 1854.72 & $-200\pm75$ & $385\pm70$ & $0.81\pm0.09$ & $0.61\pm0.05$ \\
            & Al~{\sc iii}& 1862.80 & $-230\pm70$ & $450\pm70$ & $0.78\pm0.09$ & $0.75\pm0.04$ \\
\hline
BG1501+3042	& Si {\sc ii} & 1260.42 & $-310\pm100^{\rm a)}$ & $850\pm150^{\rm a)}$ & $1.97\pm0.35^{\rm a)}$ & $0.56\pm0.10^{\rm a)}$  \\ 
            & O {\sc i} & 1302.16 & $-250\pm100^{\rm c)}$ & $725\pm125^{\rm c)}$ & $1.27\pm0.25^{\rm c)}$ & $0.64\pm0.08^{\rm c)}$ \\ 
            & Si {\sc ii} & 1304.37 & $-250\pm100^{\rm c)}$ & $725\pm125^{\rm c)}$ & $1.27\pm0.25^{\rm c)}$ & $0.64\pm0.08^{\rm c)}$ \\ 
            & C {\sc ii} & 1334.53 & $-190\pm90$ & $550\pm100$ & $1.73\pm0.40$ & $0.48\pm0.11$ \\
	    	& Si {\sc iv} & 1393.76 & $-180\pm100^{\rm a)}$ & $770\pm150^{\rm a)}$ & $2.81\pm0.45^{\rm a)}$ & $0.23\pm0.12^{\rm a)}$ \\
	    	& Si {\sc iv} & 1402.77 & $-210\pm90$ & $480\pm100$ & $1.85\pm0.35$ & $0.51\pm0.09$ \\
		    & Si {\sc ii} & 1526.70& $-140\pm90$ & $<500$ & $1.16\pm0.21$ & $0.67\pm0.15$ \\	    	
	    	& C {\sc iv} & 1548.20 & $-420\pm200^{\rm b)}$ & $1100\pm250^{\rm b)}$ & $4.5\pm0.50^{\rm b)}$ & $0.40\pm0.13^{\rm b)}$ \\
	    	& C {\sc iv} & 1550.78 & $-420\pm200^{\rm b)}$ & $1100\pm250^{\rm b)}$ & $4.5\pm0.50^{\rm b)}$ & $0.40\pm0.13^{\rm b)}$ \\ 	    	
	    	& Al {\sc ii} & 1670.78 & $-190\pm100$ & $<500$ & $0.73\pm0.25$ & $0.72\pm0.07$ \\
\hline
\end{tabular}
\end{adjustbox}
\end{center}
\textbf{Notes. ---} (1) LAE; (2) Ion; (3) Vacuum wavelength in the rest-frame; (4) Centroid velocity relative to the systemic redshift; (5) Full width half maximum corrected for the instrumental broadening; (6) Rest-frame equivalent width; and (7) Ratio between the residual intensity of the line and the continuum level. \\
$^{\rm a)}$ Affected by other absorption lines from intervening systems at lower redshifts; $^{\rm b)}$ and $^{\rm c)}$ Refers to the C~{\sc iv} 1548,1550\AA~doublet, and to the blended O~{\sc i}~1302\AA~and Si {\sc ii}~1304\AA~absorption, respectively.
\end{table*}

We fit symmetric Gaussian profiles to the low- and high-ionization ISM lines that are detected with high significance ($>5\sigma$). The strength (i.e. rest-frame equivalent widths, $EW_{0}$), the intrinsic FWHM (corrected for the instrumental broadening, see Table \ref{tab2}), and the residual intensity ($I/I_{0}$) of the ISM lines are listed in Table \ref{table_abs_line}. Note however that the measurements of $EW_{0}$ and $I/I_{0}$ in BG0742+3341 should be regarded as lower and upper limits, respectively, due to the foreground light contamination of the lens galaxy in its spectrum. Nevertheless, this effect does not change the main results of this work.
Almost all lines appear spectrally resolved, even for BG1501+3042, whose spectrum was recorded with low spectral resolution. The exception is BG0742+3341, in which the low-ionization ISM lines appear spectrally unresolved (with intrinsic $\rm FWHM < 230$~km~s$^{-1}$). 

\begin{figure*}
  \centering
  \includegraphics[width=0.67\textwidth]{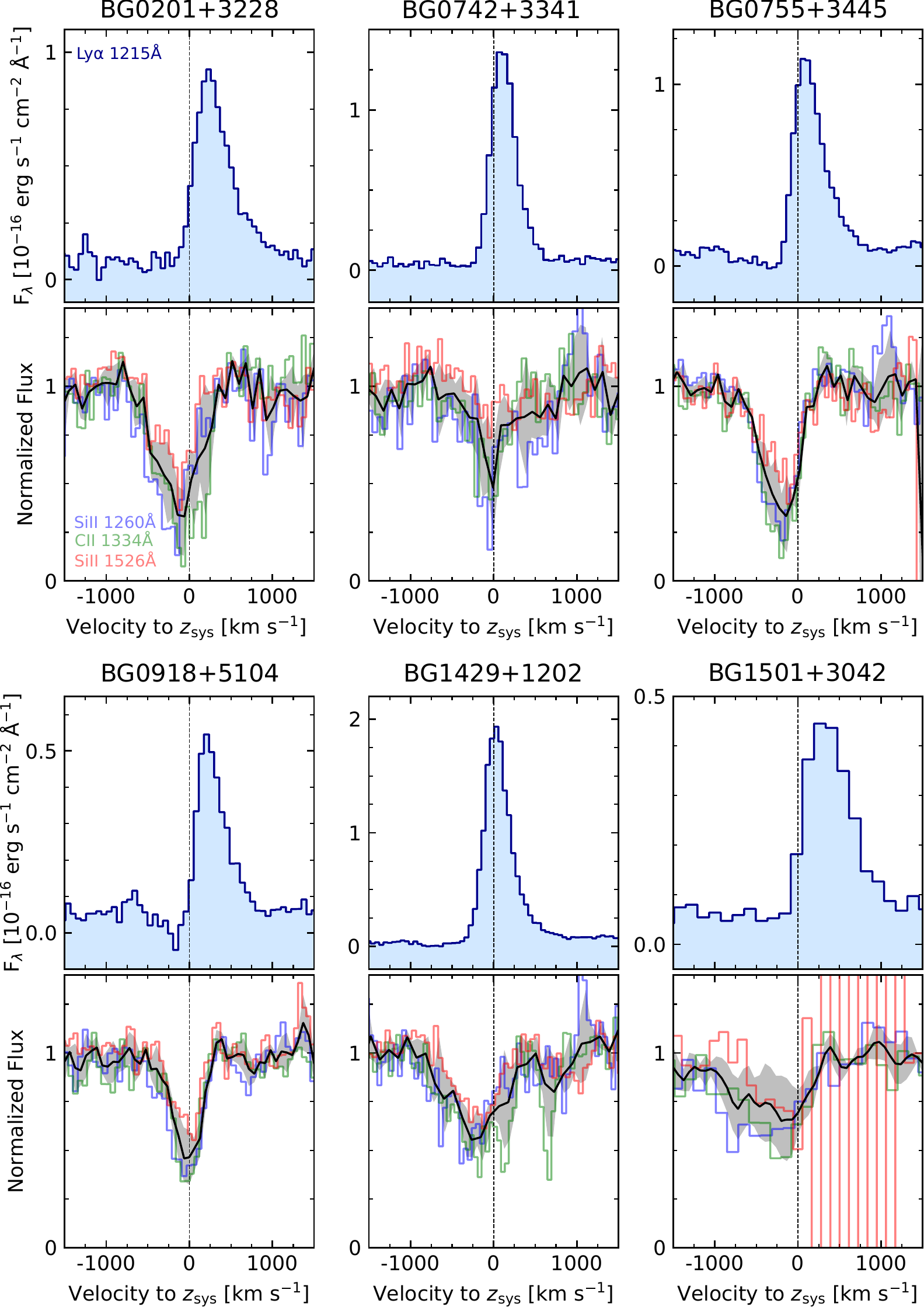}
  \caption{Ly$\alpha$ emission (upper panels) and low-ionization interstellar absorption features (lower panels; Si~{\sc ii} 1260\AA, C~{\sc ii}~1334\AA, and Si~{\sc ii}~1526\AA{ }in blue, green, and red, respectively) on the systemic velocity (measured using stellar absorption, nebular and fine-structure emission lines, see Section \ref{sys}). Ly$\alpha$ appears to be redshifted, while the interstellar absorption lines are blueshifted (average profile in black with the standard deviation in gray shadow), which is consistent with galaxy-scale outflows of material from these galaxies in the form of a wind, similar to those seen in other star-forming galaxies at $z \sim 3$ \protect\citep[e.g.,][]{shapley2003, steidel2010}. The interstellar absorption lines in the LAEs BG0742+3341 and BG0918+5104 show, however, a minimum intensity close to the systemic velocity of these galaxies.}
  \label{fig:5}
\end{figure*}

Most LAEs show ISM absorption lines blueshifted to their systemic redshifts. The velocity offset is noticeable for the most luminous LAEs, BG0201+3228, BG0755+3445, BG1429+1202, and BG1501+3042, where the ISM absorption lines appear blueshifted by $ -100 \pm 70$, $ -150 \pm 80$, $-230 \pm 60$, and $ -175 \pm 100$ km s$^{-1}$, respectively. 
On the other hand, for the less luminous LAEs, BG0742+3341 and BG0918+5104, the ISM absorption lines have the minimum intensity close to (and consistent with) the systemic velocities. Nevertheless, for these ones the spectral profiles appear slightly asymmetric with a pronounced blueshifted absorption tail. Figure \ref{fig:5} (lower panels) shows the velocity plots of several normalized low-ionization ISM absorption lines, relative to the systemic redshift of each lensed LAE measured in Section \ref{sys}.

The strength of the ISM lines also depends on the geometric covering fraction of the gas, $C_{f}$, and the ion column density, $N$. The covering fraction of an ion can be estimated from:

\begin{equation}\label{od}
    C_{f} = \frac{1 - I/I_{0}}{1-e^{- \tau}},
\end{equation}

\noindent
where $I$ is the residual intensity, $I_{0}$ the continuum level, and $\tau$ is the optical depth of an absorption line, that is related with $N$ (in units of cm$^{-2}$) by \citep{savage1991}:

\begin{equation}
    \tau = N f \lambda \frac{\pi e^{2}}{m_{e} c} = \frac{N f_{\rm osc} \lambda}{3.768\times10^{14}},
\end{equation}

\noindent
where $f_{\rm osc}$ and $\lambda$ are the ion oscillator strength \citep[][]{morton1991} and the transition wavelength (\AA). However, ISM lines in LBGs and LAEs present typically saturated profiles when observed with high spectral resolution \citep[e.g.,][]{pettini2002, des2010, jones2012, jones2013, marques2018}. 
To test if some of these lines are saturated or not, we consider the linear part of the curve of growth and use the ratios of $EW_{0}$ of different transitions of a given ion. In such case, considering the three transitions of Si~{\sc ii} detected in our spectra, at 1260\AA, 1304\AA, and 1526\AA, we would expect ratios of $EW_{0}\rm (1260\AA)$/$EW_{0}\rm (1526\AA) \simeq 5$, $EW_{0}\rm (1260\AA)$/$EW_{0}\rm (1304\AA) \simeq 10$, and $EW_{0}\rm (1304\AA)$/$EW_{0}\rm (1526\AA) \simeq 0.5$.
However, our LAEs show average values of $EW_{0} \rm (1260\AA)$/$EW_{0} \rm (1526\AA) = 1.8\pm0.9$, $EW_{0} \rm (1260$\AA)$/EW_{0}\rm (1304\AA) = 2.1\pm0.3$, and $EW_{0}\rm (1304\AA)$/$EW_{0}\rm (1526\AA) = 0.9\pm0.4$, suggesting that these transitions are likely saturated. 

We thus assume the optically thick regime, i.e. $\tau \gg 1$, so that equation \ref{od} simplifies to $C_{f} = 1 - I/I_{0}$. In order to minimize the statistical uncertainty we derive the average absorption line profile of ISM lines using the low-ionization metal lines Si~{\sc ii}~1260\AA, C~{\sc ii}~1334\AA, and Si~{\sc ii}~1526\AA. Figure \ref{fig:5} shows the normalized intensity as a function of velocity of these ISM lines. 
The estimated covering fractions using the three low-ionization metal absorption lines, $C_{f}$~(low-ISM), are between $\sim 0.3-0.7$.
For BG0742+3341, the measured $C_{f}$~(low-ISM) should be regarded as a lower limit as all low-ISM lines appear spectrally unresolved. The same may apply for BG1501+3042 whose spectrum was recorded with low spectral resolution ($\simeq 450$~km~s$^{-1}$ FWHM). Although ISM lines are spectrally resolved for this LAE, the measured values of $I/I_{0}$ are likely overestimated.

\cite{reddy2016} found that low-ionization metal absorption lines have covering fractions smaller than those from the H~{\sc i} gas. From the analysis of H~{\sc i} and low-ionization ISM lines of 18 star-forming galaxies, \cite{gazagnes2018} found that the Si~{\sc ii}~1260\AA{ }covering fraction scales linearly with the H~{\sc i} covering fraction (equation 12 in \citealt{gazagnes2018}; see also \citealt{chisholm2018}). Assuming again that the gas is optically thick, we use the values of $I/I_{0}$ of the low-ionization Si~{\sc ii}~1260\AA{ }line (see Table \ref{table_abs_line}) to measure the $C_{f}$ (Si~{\sc ii}). Following \cite{gazagnes2018}, we find $C_{f}$~(H~{\sc i}) between $\simeq 0.68-0.96$, larger than those measured using the average absorption line profile of ISM lines. All measurements are listed in Table \ref{all_laes}. In Section \ref{sec:discussion}  we discuss our measurements with those obtained for other LAEs and LBGs \citep[][]{jones2013, shibuya2014}.

\subsection{Stellar Metallicity From Stellar Population Synthesis Models}\label{age}

The most massive and hottest stars also produce strong outflows of material. The resulting spectral profile is characterized by a redshifted emission with a corresponding blueshifted absorption produced by material moving away from the stars (also called P-Cygni profile). The strength of P-Cygni profiles depends on the density of stellar winds and therefore, on the mass loss rate, which is dependent on the stellar metallicity. Because wind lines are produced by the most massive stars, P-Cygni profiles are also sensitive to the age and IMF of the stellar population. The most prominent stellar wind features in the UV are the N~{\sc v}~1238,1242\AA{ }and C~{\sc iv}~1548,1550\AA{ }high-ionization lines. 

In order to study the stellar wind features in our LAEs, we generate high-resolution ($0.4$\AA) continuum-normalized UV spectra computed with the spectral synthesis code {\sc Starburst99}\footnote{\url{http://www.stsci.edu/science/starburst99/docs/default.htm}} \citep{leitherer1999, leitherer2010}. Standard Geneva tracks with stellar metallicities ($Z_{*}/Z_{\odot}$, assuming $Z_{\odot}=0.02$) of $0.05$, $0.2$, $0.4$, $1.0$, and $2.0$ were generated. We consider models with continuous star-formation histories with a fixed age of 100 Myr and an IMF with a power slope index $\alpha = -2.35$ over the mass range $ 0.5 < M_{*}/ \rm M_{\odot} < 100$ to minimize the well-known metallicity-age-IMF degeneracy. We are aware on the limitations of these models, particularly in fixing the age of 100 Myr. Note, however, that continuous star-formation models reach equilibrium in the UV in a few tens of Myr, after which become almost time independent \citep[e.g.,][]{cabanac2008, james2014, steidel2016, marques2018}. Therefore, our models are still valid for ages $\gtrsim 30$ Myr. Considering different star formation histories (burst), and the age or IMF as free parameters would introduce large systematics in our analysis given the limited S/N of our spectra.

\begin{figure}
\centering
  \includegraphics[width=0.3\textwidth]{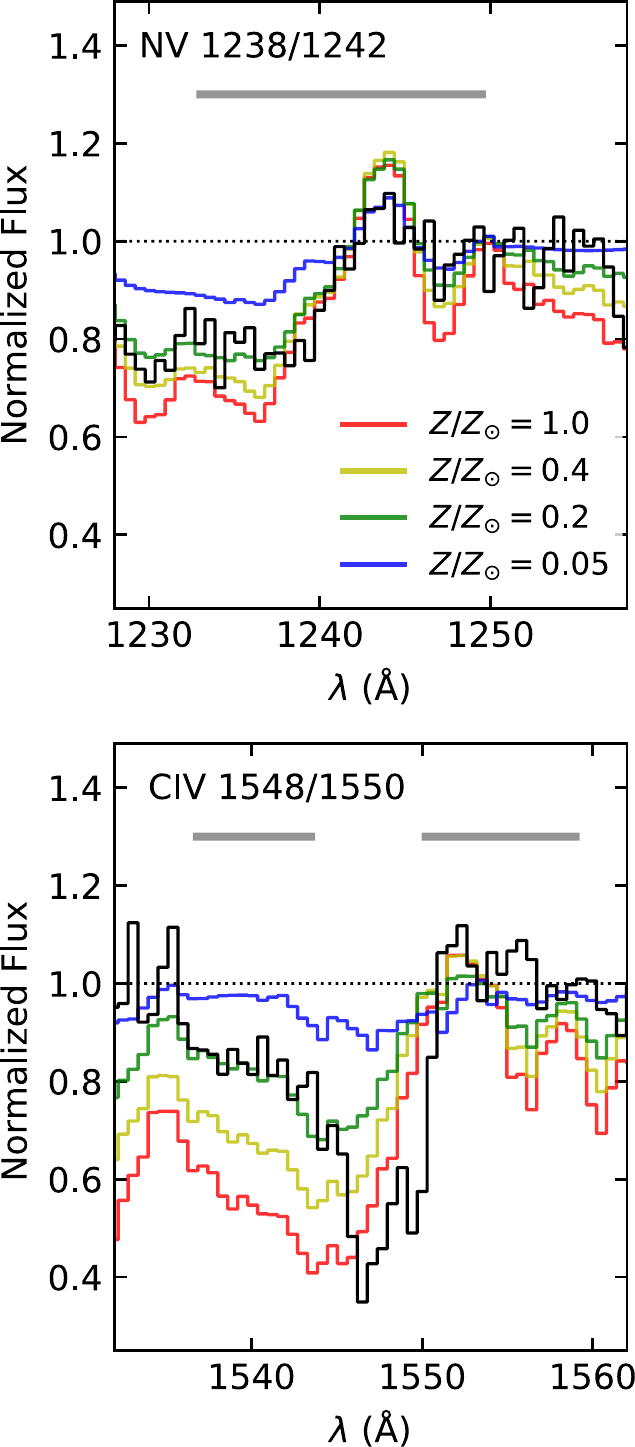}
    \caption{Comparison of the normalized composite spectrum (black) constructed from the GTC individual spectra with {\sc Starburst99} models around the wind lines N~{\sc v} (up) and C~{\sc iv} (down). {\sc Starburst99} models use a a stellar population age of $100$~Myr and metallicities of $0.05$ (blue), $0.2$ (green), $0.4$ (yellow), and $1.0 Z_{\odot}$ (red), as indicated. The spectral windows used to perform the $\chi^{2}$ minimization are marked in gray. Regions of strong interstellar absorption in C~{\sc iv} were excluded in the fit. Models with metallicity $Z_{*} / Z_{\odot} = 0.2$ reproduce quite well the line profiles of the high-ionization N~{\sc v} and C~{\sc iv} lines, in particular the blueshifted absorption of the P-Cygni profile of C~{\sc iv} that is the most metallicity sensitive feature in the UV \citep{chisholm2019}.
    \label{fig:starburst99}}
\end{figure}

The analysis is performed in the wind N~{\sc v} and C~{\sc iv} lines, those that show prominent P-Cygni profiles. Our GTC spectra do not show any clear wind features in other lines that are conspicuous in the spectra of individual O stars (O~{\sc v}~1371\AA, Si~{\sc iv}~1383,1402\AA, or N~{\sc iv}~1719\AA), suggesting that the more numerous B stars are diluting the contribution of the O stars. 
{\sc Starburst99} outputs were smoothed to the spectral resolution of our GTC spectra and rebined to a spectral element of $0.58$ \AA~pix$^{-1}$. 
The GTC spectra were firstly normalized using local continuum windows that are free of strong absorption and emission features. 
Large spectral windows were used in the fit (marked in gray in Figure \ref{fig:starburst99}) to account for stellar winds with terminal velocities exceeding $\sim 2000$~km~s$^{-1}$ \citep[e.g.,][]{pettini2000, marques2018}, from $1233-1250$ \AA{ }for N~{\sc v} and $1537-1559$ \AA{ }for C~{\sc iv}. We exclude in the fit spectral regions affected by strong sky-subtracted residuals and strong interstellar absorption (in C~{\sc iv}, from $1539-1552$ \AA) as well as other absorption lines produced by intervening systems at lower redshifts (see Appendix \ref{low-z}). We exclude BG0742+3341 in this analysis, due to the low S/N in the continuum and the light contamination from the foreground lens galaxy. 

The total $\chi^{2}$ is computed by comparing all the included spectral elements of our normalized spectra with those from the models with different metallicities:

\begin{equation}\label{chi2}
    \chi^{2} = \sum_{i} \left( \frac{\textrm{model}(i) - \textrm{data}(i)}{\sigma (i)} \right)^{2}, 
\end{equation}

\noindent
where $\sigma (i)$ is the error of the spectral element $i$. Table \ref{SB99_table} shows the values of the metallicites that minimized $\chi^{2}$ for a given synthesis model. We also show the significance of the difference between the goodness of fit of each model and the best-fitting model, $\Delta \sigma$, calculated for a $\chi^{2}$ distribution with certain number of degrees of freedom (d.o.f).

\begin{table}
\begin{center}
\caption{ $\chi^{2}$ minimization for stellar models. \label{SB99_table}}
\begin{tabular}{l c c c c c}
\hline \hline
\smallskip
\smallskip
LAE &  d.o.f & $Z_{*}/Z_{\odot}$ & $\chi^{2}$ & $\chi^2$/d.o.f &  $\Delta \sigma$\\
  & (1) & (2) & (3) & (4) & (5)\\
\hline 
BG0201+3228  &  47  & 0.05 & 67.12  &  1.43  &  1.34 \\
             &      & 0.20 & 54.18  &  1.15  &  0.00 \\
             &      & 0.40 & 69.75  &  1.49  &  1.61 \\
             &      & 1.00 & 108.43 &  2.31  &  5.60 \\
\hline
BG0755+3445  &  54  & 0.05 & 85.03  &  1.58  &  2.21 \\
             &      & 0.20 & 62.09  &  1.15  &  0.00 \\
             &      & 0.40 & 79.73  &  1.48  &  1.70 \\
             &      & 1.00 & 126.19 &  2.33  &  6.17 \\
\hline
BG0918+5104  &  54  & 0.05 & 61.68  &  1.14  &  0.37 \\
             &      & 0.20 & 57.79  &  1.07  &  0.00 \\
             &      & 0.40 & 99.90  &  1.85  &  4.05 \\
             &      & 1.00 & 183.90 &  3.41  &  12.14 \\
\hline
BG1429+1202  &  42  & 0.05 & 95.36  &  2.27  &  4.72 \\
             &      & 0.20 & 57.94  &  1.38  &  0.63 \\
             &      & 0.40 & 52.14  &  1.24  &  0.00 \\
             &      & 1.00 & 64.96  &  1.55  &  1.40 \\
\hline
BG1501+3042  &  44  & 0.05 & 70.02  &  1.87  &  2.38 \\
             &      & 0.20 & 47.67  &  1.08  &  0.00 \\
             &      & 0.40 & 66.39  &  1.51  &  2.00 \\
             &      & 1.00 & 119.98 &  2.73  &  7.71 \\
\hline
\end{tabular}
\end{center}
\textbf{Notes. ---} (1) degrees of freedom; (2) metallicity of the synthesis models (assuming $Z_{\odot} = 0.02$); 
(3) total $\chi^{2}$ for each model (equation \ref{chi2}); (4) reduced $\chi^{2}$; and (5) 1-$\sigma$ deviation between the goodness of fit of each model and the best-fitting model.
\end{table}

As shown in Table \ref{SB99_table}, stellar synthesis models with metallicity $Z_{*} / Z_{\odot} \simeq 0.2$ are preferred over other solutions in almost all LAEs, with $\chi^{2}/$d.o.f ranging from $1.07$ to $1.38$. We double-check this by generating a high S/N composite spectrum constructed from the five GTC individual spectra. The resulting composite spectrum has higher S/N in the continuum and is less affected by sky-subtracted residuals and spurious features due to intervening absorption systems.
Figure \ref{fig:starburst99} shows the stacked spectrum around the wind lines N~{\sc v} and C~{\sc iv}, as well as the {\sc Starburst99} models with different metallicities. As clearly seen in this figure, models with metallicity $Z_{*} / Z_{\odot} \simeq 0.2$ reproduce quite well the line profiles in these regions, in line with the results from our $\chi^{2}$ minimization.

Note however that for two LAEs, BG0918-5104 and BG1429+1202, other solutions are also accepted with a 1-$\sigma$ deviation from the best-fit model below 1 (see Table \ref{SB99_table}). For BG0918+5104, both models with $Z_{*} / Z_{\odot} = 0.05$ and $0.20$ provide good fits to the GTC spectrum with $\rm \chi^{2}/d.o.f. = 1.14$ and $1.07$, respectively, and a $\Delta \sigma$ between these two models of only 0.37. This suggests that the metallicity of BG0918+5104 should lie between these two values, being likely the lowest-metallicity LAE in our sample, which in turn might explain the strong nebular emission seen in its spectrum (see Figure \ref{fig:fs}). On the other hand, the spectrum of BG1429+1202 is better explained by a model with stellar metallicity of $Z_{*} / Z_{\odot} \simeq 0.4$. It is worth noting that a slightly large metallicity of $Z_{*} / Z_{\odot} \sim 0.6$ was inferred recently by \cite{chisholm2019} for BG1429+1202. These authors use, however, multiple single age and metallicity models to fit the UV stellar continuum, while we are using continuous star-formation models with a fixed age of 100~Myr. Our analysis is therefore limited to discrete models with a set of five metallicities. Consequently, we can only say that a model with stellar metallicity of $Z_{*} / Z_{\odot} \simeq 0.4$ is favoured with respected to the other models ($Z_{*} / Z_{\odot} = 0.05$, $0.20$, $1.00$ or $2.00$). In addition, the blueshifted absorption of C~{\sc iv} in BG1429+1202 is severely affected by strong sky residuals (at $\lambda_{\rm obs} \sim 5890-5896$~\AA), and therefore was not included in the fit. As already shown in several works (e.g., \citealp{quider2009, quider2010, james2014, marques2018}, but in particular in \citealp{chisholm2019}) the spectral region encompassing the blueshifted P-Cygni absorption is the most metallicity sensitive feature in the UV.

\subsection{Ly$\alpha$ Line}\label{sec:lya}

Our spectra show prominent and narrow Ly$\alpha$ emission lines (see Figure \ref{fig:5}). We measure Ly$\alpha$ fluxes ranging from $(3.0 - 11.5) \times 10^{-16}$~erg~s$^{-1}$~cm$^{-2}$, much higher (by factors of $1.4$ to $4.7$) than the ones measured within the $1^{\prime \prime}$-radius BOSS fibers \citep{shu2016a}, as a consequence of the large image separation in these LAEs. By determining the stellar continuum on both sides of the Ly$\alpha$ line, we measure Ly$\alpha$ rest-frame equivalent widths $EW_{0} \simeq 16 - 50$ \AA. The Ly$\alpha$ $EW_{0}$ of BG0918+5104 and BG1501+3042 are smaller than the lower limit $EW_{0} > 20$~\AA{ } commonly adopted to define them as LAEs. In all LAEs the Ly$\alpha$ line is spectrally resolved with $\rm FWHM$ ranging between $\simeq 230 - 400$ km s$^{-1}$, by fitting a Gaussian and correcting it for the instrumental broadening. All quantities are summarized in Table \ref{all_laes}.

All LAEs show asymmetric Ly$\alpha$ line profiles (see Figure \ref{fig:5}). At our spectral resolution and depth we do not detect any clear blueshifted component emission in any LAE. For three LAEs (BG0742+3341, BG0755+3445, and BG1429+1202), however, the Ly$\alpha$ line has its maximum intensity very close to, or even consistent with the systemic velocity (see Section \ref{sys}). These are the ones that show narrower Ly$\alpha$ profiles (230-330~km~s$^{-1}$ FWHM) suggestive of less scatter from low H~{\sc i} column densities and/or that Ly$\alpha$ photons freely escape through a clumpy ISM \citep[e.g.,][]{zheng2014, verhamme2015, verhamme2018, claeyssens2019}. 
On the other hand, in addition to the prominent Ly$\alpha$ emission line, BG0201+3228 shows also an underlying broad Ly$\alpha$ absorption (see Figure \ref{all_laes}). This may indicate that the neutral gas in this LAE is inhomogeneous with both high and low column density H~{\sc i} regions \citep[see:][]{mcKinney2019, jaskot2019}, in line with the large covering fraction measured in Section \ref{ism} ($C_{f}$~(H~{\sc i})~$ \simeq 0.96$).

Our GTC spectra are subject to variable slit-losses, since the observations were carried out through narrow slits ($1.0^{\prime \prime}$ and $1.2^{\prime \prime}$ wide) covering only a fraction of the total flux of these LAEs. 
In order to put our spectra on an absolute flux scale, we use the {\sc Python} package {\sc pyphot 1.0.1}\footnote{\url{http://mfouesneau.github.io/docs/pyphot/}} to compute synthetic photometry of the GTC extracted spectra and compare to the total photometry derived in Section \ref{sec:phot} and listed in Table \ref{all_laes}.
The comparison is done using the $g$ band ($\lambda_{\rm min} \simeq 4000$ \AA, $\lambda_{\rm max} \simeq 5350$ \AA) to avoid the contamination in the LAEs spectra of the stellar light redward the rest-frame $4000$ \AA~\space break of the foreground lens, which is redshifted to the optical $i$-band in all LAE spectra. Depending on each LAE, we find correction factors to account for slit-losses ranging between $0.35$ to $0.90$.

Having the GTC spectra on an absolute flux scale, we derive the intrinsic Ly$\alpha$ luminosity of these LAEs by using the following formula:

\begin{equation}
L_{\rm Ly \alpha} = 4 \pi D_{\rm L}^{2} F_{\rm Ly \alpha}^{\rm obs} \times \frac{1}{\mu} \times \frac{1}{f},
\end{equation}

\noindent
where $L_{\rm Ly \alpha}$ is given in units of erg s$^{-1}$, $D_{\rm L}$ in cm, and $f$ is the correction factor to account for slit-losses. 

Assuming case-B recombination and the \cite{kennicutt1998} conversion, the Ly$\alpha$ luminosity can be translated into a star formation rate: 

\begin{equation}\label{eq_lya}
\rm SFR~(Ly\alpha) = 9.1 \times 10^{-43} L_{\rm Ly \alpha} \times 10^{0.4 \times A_{\rm Ly\alpha}} \times \frac{1}{1.8},  
\end{equation}

\noindent
assuming the \cite{chabrier2003} IMF, where $A_{\rm Ly\alpha} = 10.2 \times E(B-V)$ is the attenuation at $1215$\AA{ }\citep{reddy2016a}. The nebular color excess is assumed to be the same as the one measured for the stellar continuum \citep[e.g.,][]{erb2006b, puglisi2016}. 

\begin{figure*}
  \centering
  \includegraphics[width=0.70\textwidth]{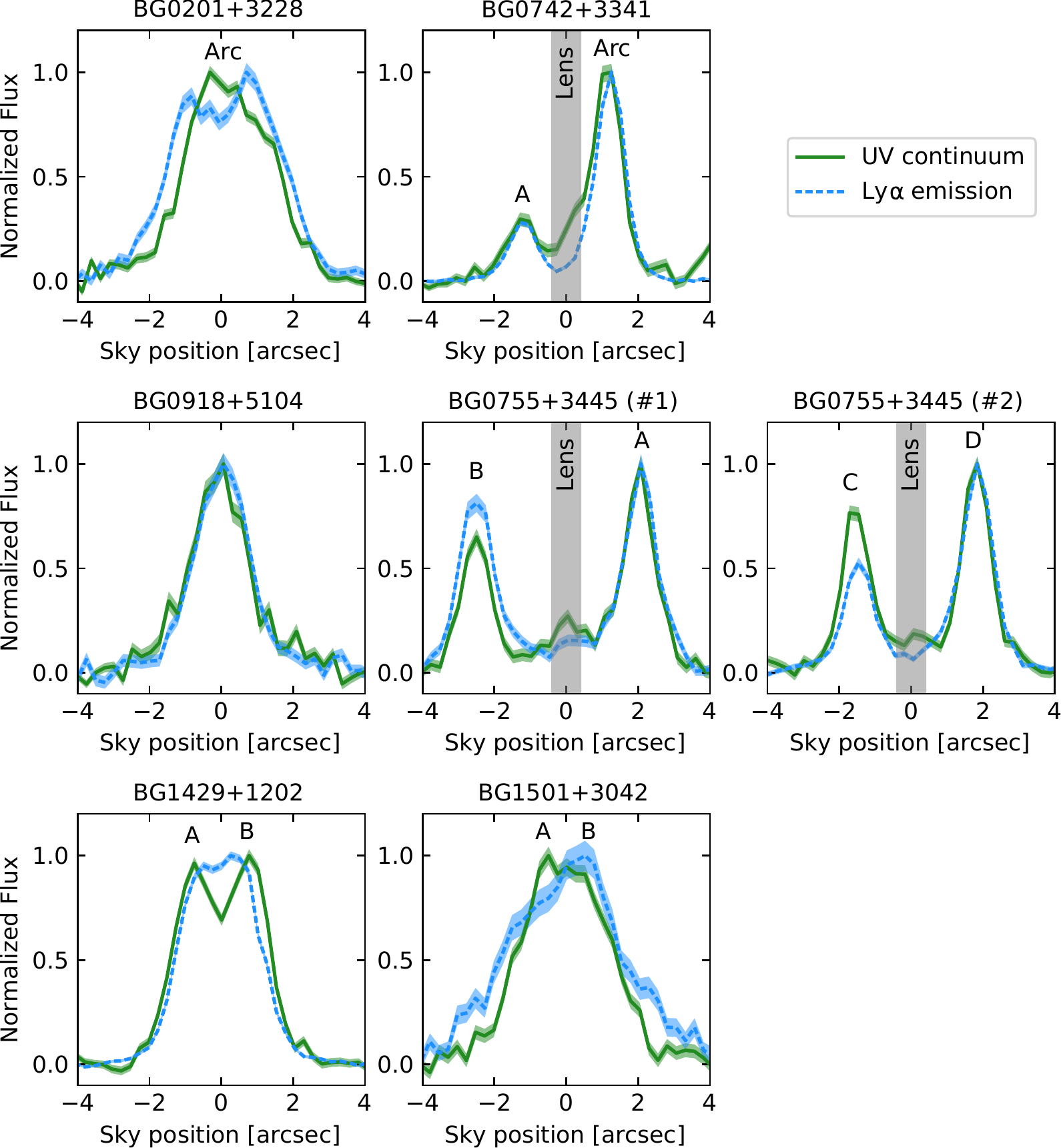}
  \caption{Normalized spatial profiles along the slit of the rest-frame UV continuum (green) and Ly$\alpha$ emission (blue). Shaded regions show the $\pm 1\sigma$ errors of the spatial profiles. For the LAEs BG0201+3228, BG1429+1202 and BG1501+3042 the peak of the Ly$\alpha$ emission is spatially offset with respect to the peak of the UV continuum. The lensed images B and C of BG0755+3445 show different Ly$\alpha$ to UV continuum ratios compared with the lensed images A and D, resulting in different $EW_{0}$ (Ly$\alpha$).}
  \label{fig:2lya}
\end{figure*}

Comparing the measurements of SFRs from the rest-frame UV (Section \ref{sfr_uv}) and Ly$\alpha$, we find Ly$\alpha$ escape fractions, defined as $f_{\rm esc}~\rm(Ly\alpha) = \rm SFR_{\rm Ly \alpha} / \rm SFR_{\rm UV}$, between $0.11$ and $0.37$ (summarized in Table \ref{all_laes}).

Note, however, that our spectroscopic  measurements of $EW_{0}$~(Ly$\alpha$) and $f_{\rm esc}$~(Ly$\alpha$) are likely underestimated, at least compared to those measured from narrow-band imaging surveys. 
Star-forming galaxies, such as LAEs or LBGs, show Ly$\alpha$ emission more spatially extended than the rest-frame UV continuum  \citep[e.g.,][]{steidel2011, momose2014, matthee2016, wisotzki2016, leclercq2017}, and in some cases with its emission spatially offset from the peak of their UV continuum \citep[e.g.,][]{shibuya2014b, matthee2016, erb2019, hoag2019}. As a result, it has been shown that photometric measurements of $EW_{0}$~(Ly$\alpha$) are on average $\sim 5$ times larger than those derived using long-slit spectra \citep[see:][]{steidel2011, erb2014}. 

This effect should be even more prominent in gravitational lensed systems, like the BELLS GALLERY LAEs. 
To investigate this, the spatial profiles of Ly$\alpha$ and UV continua of these LAEs were extracted along the slit direction by summing the spectra in specific wavelength ranges along the dispersion direction: $\approx 1210 - 1224$~\AA~\space rest-frame for the Ly$\alpha$ line, and several tens Angstroms redward and blueward of Ly$\alpha$ for the UV continuum. Both Ly$\alpha$ and UV continuum profiles are then normalized to their maximum. Figure \ref{fig:2lya} shows normalized spatial profiles of the UV continuum and Ly$\alpha$ line.
For three lensed LAEs, BG0201+3228, BG1429+1202, and BG1501+3042, we find clear differences between the Ly$\alpha$ and UV continuum 1D spatial profiles. 
The Ly$\alpha$ emission in the lensed LAEs BG0201+3228 and BG1501+3042 presents a more extended profile than the UV continuum, while for BG1429+1202 the Ly$\alpha$ emission appears brighter in the inner region between the lensed images A and B than its UV continuum \citep[similar results found in][]{marques2017}. 
It is worth noting that the Ly$\alpha$ and UV continuum emission of the lensed images B and C in BG0755+3445 show different Ly$\alpha$ to UV continuum ratios compared with the lensed images A and D, resulting in different $EW_{0}$ (Ly$\alpha$). 
Finally, the spatial distribution of Ly$\alpha$ in BG0742+3341 and BG0918+5104 appears to follow the UV continuum, although we stress that the orientation of the long slit in BG0742+3341 is not ideal to look for extended Ly$\alpha$ emission, since it was placed perpendicular to the extended arc. 

Overall, our results are insufficient for definitive conclusions on the intrinsic morphology and extension of Ly$\alpha$. However, the differences between the 1D spatial profiles of Ly$\alpha$ and UV seen in Figure \ref{fig:2lya} already suggests intrinsic differences on the morphology and/or spatial offsets between Ly$\alpha$ and UV emission. 
As a result, one should expect that the Ly$\alpha$ emission measured in the spectra is subject to larger slit-losses than the UV continuum, since the GTC long slits were centered on the brightest UV lensed images of each LAE without any information on the light distribution of the Ly$\alpha$ emission,  
Different morphologies and/or spatial offsets in the source plane can also lead to differential lensing magnification between the UV and Ly$\alpha$ emission. As Ly$\alpha$ emission is found to be statistically more extended than the UV continuum in star-forming galaxies, the total magnification of the Ly$\alpha$ emission in our LAEs is likely to be lower than that of the more compact UV continuum. 

Integral field spectroscopy or narrow-band imaging observations of the Ly$\alpha$ emission are required to better constrain the intrinsic properties of Ly$\alpha$ (e.g, morphology, luminosity, escape fraction and rest-frame equivalent width) and investigate possible differential magnification between the Ly$\alpha$ and UV emission.

\subsection{Other Nebular lines}

In addition to the Ly$\alpha$ line, we also detect other nebular emission lines, such as O~{\sc iii}] and C~{\sc iii}]. Prominent emission in these lines are found predominantly in BG0755+3445 and BG0918+5104 and are detected with high significance (see Figure \ref{fig:fs}), whereas for the remaining LAEs these lines are only barely detected (even at $<3 \sigma$) or fall in regions of strong night sky lines. Despite this, we provide measurements and upper limits of their fluxes and equivalent widths. These measurements will be useful to gain insight on the nature of the ionizing radiation field of our LAEs and will be investigated further in detail in Section \ref{sfr_agn}.

We fit symmetric Gaussian profiles to measure the flux and equivalent width of these lines. For C~{\sc iii}] we use two Gaussian profiles to fit simultaneously the C~{\sc iii}]~1906,1908\AA{ }doublet, except for BG1429+1202 and BG1501+3042 where we use a single profile due to the low S/N. All nebular lines appear spectrally unresolved at our resolution, so that their profiles have $\rm FWHM<200$~km~s$^{-1}$ ($\rm FWHM<450$~km~s$^{-1}$ for BG1501+3042). For non-detections, we provide $3\sigma$ upper limits, assuming spectrally unresolved profiles. Some examples of the fitting are shown in Figure \ref{fig:fs} (blue dashed-lines). 

These measurements are listed in Table \ref{tab:emission}. We also provide measurements or limits ($3\sigma$) of the intrinsic luminosities of these lines, i.e. corrected for magnification and slit-losses, that can be useful for future investigation of the physical properties of the ionized gas.

\begin{table}
\begin{center}
\caption{Nebular emission lines.}\label{tab:emission}
\begin{tabular}{lcccc}
\hline  \hline 
Line & $\lambda_{\rm rest}$   & $EW_{0}$  & Flux & Luminosity \\  
     &  (1)  &  (2) & (3)  &  (4) \\
\hline
\multicolumn{5}{c}{BG0201+3228}\\ 
\hline 
He~{\sc ii} & 1640.42  & $<0.10$        & $<1.61$       & $<0.25$\\
{}[O~{\sc iii}] & 1660.81  & $0.08\pm0.06$  & $0.45\pm0.32$ & $0.07\pm0.05$\\
{}O~{\sc iii}] & 1666.15  & $0.19\pm0.11$  & $0.56\pm0.32$ & $0.09\pm0.05$\\
{}[C~{\sc iii}] & 1906.68  & $0.28\pm0.20$  & $1.23\pm0.72$ & $0.19\pm0.11$\\
{}C~{\sc iii}] & 1908.68  & $0.30\pm0.15$  & $1.34\pm0.54$ & $0.21\pm0.09$\\
\hline 
\multicolumn{5}{c}{BG0755+3445}\\ 
\hline
He~{\sc ii} & 1640.42  & $<0.31$       & $<1.27$       & $<0.16$\\
{}[O~{\sc iii}] & 1660.81  & $<0.30$       & $<1.10$       & $<0.14$\\
{}O~{\sc iii}] & 1666.15  & $0.33\pm0.25$ & $0.25\pm0.19$ & $0.03\pm0.02$\\
{}[C~{\sc iii}] & 1906.68  & $1.23\pm0.30$ & $1.04\pm0.25$ & $0.13\pm0.03$\\
{}C~{\sc iii}] & 1908.68  & $1.10\pm0.28$ & $0.92\pm0.23$ & $0.12\pm0.03$\\
\hline
\multicolumn{5}{c}{BG0918+5104}\\ 
\hline
He~{\sc ii} & 1640.42  & $<5.01$       &  $<8.36$  & $<0.99$\\
{}[O~{\sc iii}] & 1660.81  & $0.36\pm0.16$ & $0.56\pm0.23$ & $0.07\pm0.03$\\
{}O~{\sc iii}] & 1666.15  & $0.84\pm0.16$ & $1.29\pm0.24$ & $0.15\pm0.03$\\
{}[C~{\sc iii}] & 1906.68  & --  & --  & --   \\
{}C~{\sc iii}] & 1908.68  & $1.51\pm0.21$ & $2.51\pm0.33$ & $0.30\pm0.04$\\
\hline 
\multicolumn{5}{c}{BG1429+1202}\\ 
\hline
He~{\sc ii} & 1640.42  & $<0.48$        &  $<1.92$  & $<1.34$\\
{}[O~{\sc iii}] & 1660.81  & $<0.30$        &  $<1.21$  & $<0.84$\\
{}O~{\sc iii}] & 1666.15  & $0.17\pm0.12$  & $0.66\pm0.47$ & $0.46\pm0.33$\\
{}[C~{\sc iii}] & 1906.68  & $0.83\pm0.40^{\rm a}$  & $2.39\pm1.13^{\rm a}$ & $1.67\pm0.79^{\rm a}$\\
{}C~{\sc iii}] & 1908.68  & $0.83\pm0.40^{\rm a}$  & $2.39\pm1.13^{\rm a}$ & $1.67\pm0.79^{\rm a}$\\
\hline 
\multicolumn{5}{c}{BG1502+3042}\\ 
\hline
He~{\sc ii} & 1640.42  & $<1.55$ &  $<2.41$  & $<2.19$\\
{}[O~{\sc iii}] & 1660.81  & $<1.55$ &  $<2.41$  & $<2.19$\\
{}O~{\sc iii}] & 1666.15  & $<1.55$ &  $<2.41$  & $<2.19$  \\
{}[C~{\sc iii}] & 1906.68  & $0.46\pm0.24^{\rm a}$  & $0.67\pm0.35^{\rm a}$ & $0.61\pm0.32^{\rm a}$\\
{}C~{\sc iii}] & 1908.68  & $0.46\pm0.24^{\rm a}$  & $0.67\pm0.35^{\rm a}$ & $0.61\pm0.32^{\rm a}$\\
\hline
\end{tabular}
\end{center}
\textbf{Notes. ---} (1) vacuum wavelength in the rest-frame; (2) rest-frame equivalent width; 
(3) observed flux in units of $10^{-17}$~erg~s$^{-1}$~cm$^{-2}$. For non-detections upper limits refer to $3\sigma$; (4) intrinsic luminosity (corrected for magnification and slit-losses) in units of $10^{8} \times L_{\odot}$ (where $L_{\odot} = 3.826 \times 10^{33}$~erg~s$^{-1}$).\\
$^{\rm a}$ Refer to the measurements of the C~{\sc iii}]~1906,1908\AA{ }doublet using a single Gaussian profile.  
\end{table}

\section{Discussion}\label{sec:discussion}

\subsection{The origin of large $EW_{0}~ \rm (Ly\alpha$) and $f_{\rm esc}~ \rm (Ly\alpha$)}\label{sec:6.1}

Our sample selection consisted of the follow-up of UV-bright BELLS GALLERY lensed galaxies in order to achieve high S/N continuum spectra. Despite this, these galaxies show prominent Ly$\alpha$ emission lines with $EW_{0}$ up to $\simeq 50$~\AA, even noting that these measurements are likely underestimated (see Section \ref{sec:lya}). 
We will now compare other properties of the BELLS GALLERY LAEs to those from other Ly$\alpha$- and UV-selected galaxies.

\begin{figure*}
  \centering
  \includegraphics[width=0.96\textwidth]{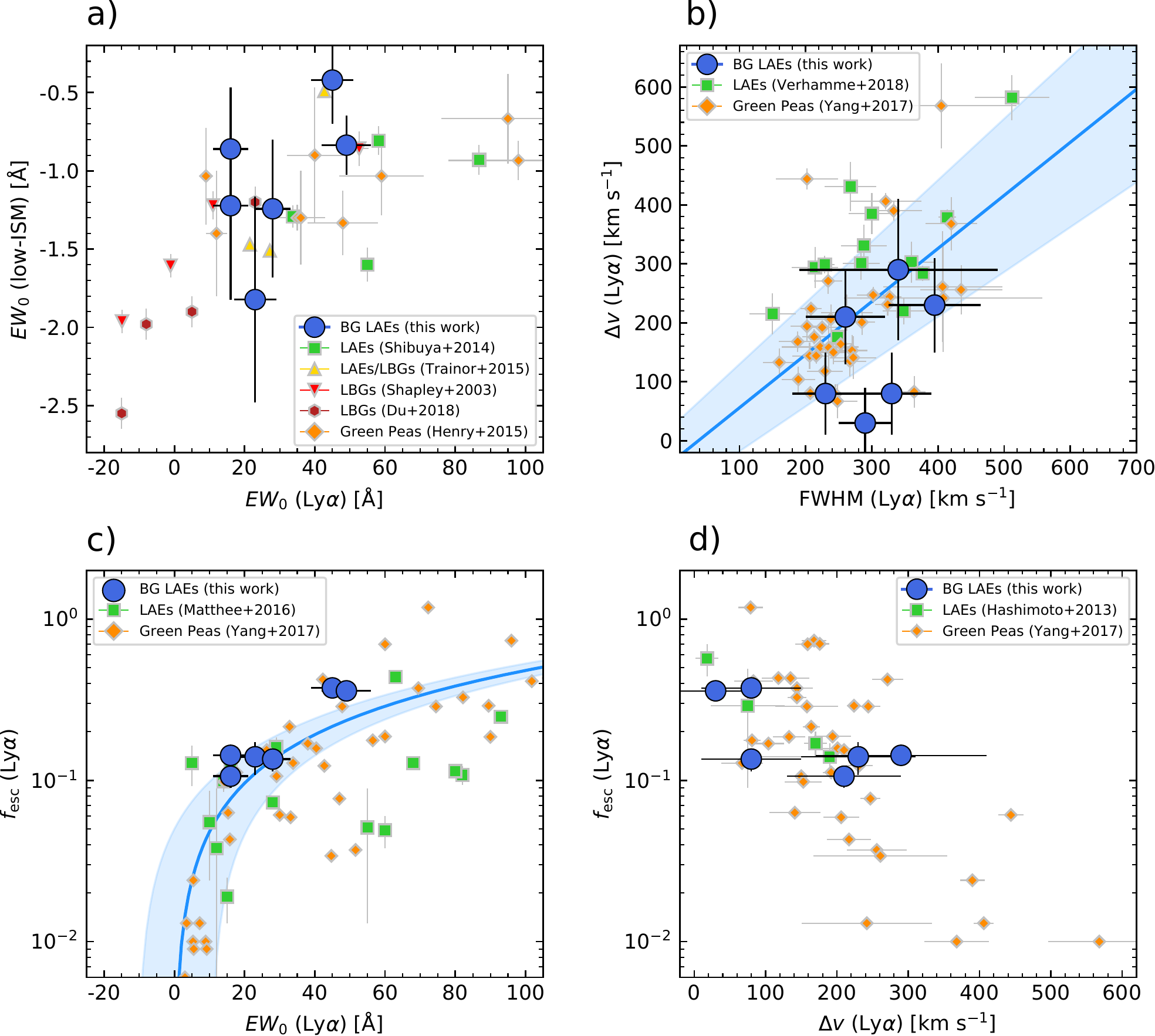}
  \caption{Comparison of several properties of the BELLS GALLERY LAEs (blue circles) to those from other LAEs \protect\citep[][]{hashimoto2013, shibuya2014, trainor2015,matthee2016, verhamme2018}, LBGs \protect\citep[][]{shapley2003, trainor2015, du2018}, and nearby high-redshift analogs \protect\citep[Green Pea galaxies][]{henry2015, yang2017b}: a) rest-frame equivalent width of the averaged low-ionization ISM lines, $EW_{0} \rm (low-ISM)$, as a function of Ly$\alpha$ equivalent width, $EW_{0} \rm (Ly\alpha)$; b) relation between the velocity offset of the Ly$\alpha$ peak relative to the systemic redshift, $\Delta v \rm (Ly\alpha)$, as a function of the FWHM of the Ly$\alpha$. These two quantities are correlated \protect\citep[blue line from][]{verhamme2018} and depend mostly on the H~{\sc i} opacity; c) Ly$\alpha$ escape fraction, $f_{\rm esc} \rm (Ly\alpha$), as a function of $EW_{0} \rm (Ly\alpha)$. Blue line is the empirical relation derived by \protect\cite{sobral2019}; and d) $f_{\rm esc} \rm (Ly\alpha$) as a function of $\Delta v \rm (Ly\alpha)$. In general, large $EW_{0}$~(Ly$\alpha$) and $f_{\rm esc}$~(Ly$\alpha$) are associated with both weak low-ionization ISM absorption lines, narrow Ly$\alpha$ line profiles and small $\Delta v  \rm (Ly\alpha)$.}
  \label{fig:last}
\end{figure*}

In the top left panel of Figure \ref{fig:last} we show the relation between the rest-frame equivalent widths of the averaged low-ionization ISM lines (using the Si~{\sc ii}~1260\AA, C~{\sc ii}~1334\AA{ }and Si~{\sc ii}~1526\AA{ }absorption lines) and the Ly$\alpha$ emission of our LAEs. We also add to this plot the measurements of $EW_{0}$'s of low-ionization ISM lines and Ly$\alpha$ of other $z =2 - 4$ LAEs and LBGs \citep[][]{shapley2003, shibuya2014, trainor2015, du2018}, as well as other nearby high-redshift analogs \citep[Green Pea galaxies;][]{henry2015}.  

Our measurements confirm previous findings that weaker (i.e. lower $EW_{0}$) low-ionization ISM absorption lines are associated with stronger Ly$\alpha$ emission \citep[e.g.,][]{shapley2003, shibuya2014, du2018}. Such relation suggests that strong LAEs show, on average, weaker low-ionization ISM absorption lines than typical LBGs due to their lower covering fractions and low column densities that allows Ly$\alpha$ photons to escape more efficiently \citep[e.g.,][]{erb2010, jaskot2014}.

While the covering fraction is known to be sensitive to the spectral resolution, the physical and geometric properties of the ISM can also be  investigated through the strength and spectral shape of the observed Ly$\alpha$ line. 

Our LAEs show relatively narrow Ly$\alpha$ profiles, with intrinsic (i.e. corrected for the instrumental broadening) FWHMs around $230-400$~km~s$^{-1}$, suggestive of a reduced number of scatterings in the ISM with a low H~{\sc i} column density \citep[e.g.,][]{verhamme2015}. We also find small velocity offsets (redshifted) of the Ly$\alpha$ peak relative to the systemic redshifts with a mean value and scatter $\Delta v~\rm (Ly\alpha) = 150\pm 90$~km~s$^{-1}$. This is similar to the values measured in typical LAEs \citep[$\Delta v~\rm (Ly\alpha) \sim 100-300$~km~s$^{-1}$, e.g.,][]{finkelstein2011a, hashimoto2013, shibuya2014, hashimoto2015, trainor2015, amorin2017, verhamme2018}, but lower than those found in UV-selected LBGs, by a factor of $\approx 2.7$ ($\Delta v~\rm (Ly\alpha) \sim 400$~km~s$^{-1}$; e.g., \citealt{steidel2010}). In particular, BG0742+3341, BG0755+3445 and BG1429+1202 show very small values of $\Delta v~\rm (Ly\alpha)$, around $30-80$~km~s$^{-1}$, that are expected for objects with density-bounded H~{\sc ii} regions, so that Ly$\alpha$ photons would escape relatively unimpeded from the ionizing sources \citep[e.g.,][]{verhamme2015}. 

In Figure \ref{fig:last} (top right panel) we show the relation between these two quantities derived in our LAEs, $\Delta v~\rm (Ly\alpha)$ and FHWM~(Ly$\alpha$). We compare our results with those found in other high-$z$ LAEs and local analogs. As shown already by \cite{verhamme2015, verhamme2018} these two quantities are tightly correlated (blue line in Figure \ref{fig:last}) and depend mostly on the H~{\sc i} opacity. This clearly suggests that the BELLS GALLERY LAEs studied here, in particular those showing small values of $\Delta v~\rm (Ly\alpha)$ and FHWM~(Ly$\alpha$), likely present low H~{\sc i} column densities.  

This scenario is further supported by the relation between the Ly$\alpha$ escape fraction and $EW_{0}$~(Ly$\alpha$) and $\Delta v~\rm (Ly\alpha)$ shown in the bottom panels of Figure \ref{fig:last}. 
Our BELLS GALLERY LAEs show Ly$\alpha$ escape fraction ranging between 0.11 and 0.37 (see Table \ref{all_laes}), with a mean value and scatter $f_{\rm esc}$~(Ly$\alpha$)$ = 0.21\pm0.11$, in between the range of values found for $z=2-4$ UV-selected galaxies ($f_{\rm esc}$~(Ly$\alpha$)~$\approx 0.05$; e.g., \citealt{kornei2010, cassata2015}) and other strong Ly$\alpha$ emitters ($f_{\rm esc}$~(Ly$\alpha$)~$\approx 0.30$; e.g., \citealt{nakajima2012, hashimoto2013, wardlow2014, trainor2015, matthee2016}). We recover previous findings that large Ly$\alpha$ escape fraction is associated with small $\Delta v~\rm (Ly\alpha)$ \citep{hashimoto2013, verhamme2015, yang2017b, verhamme2018} and strong $EW_{0}$'s~(Ly$\alpha$) \citep[e.g.,][]{trainor2015, matthee2016, sobral2017, yang2017b, izotov2019, mcKinney2019}, in line with the empirical relation derived by \cite{sobral2019}. 

Altogether, our findings support the idea that the Ly$\alpha$ emissivity and escape fraction are anti-correlated with the strength of the ISM absorption lines, and the broadness and the velocity offset of the Ly$\alpha$ line. As an example, the two stronger LAEs in our sample, BG0742+3341 and BG1429+1202, those with larger Ly$\alpha$ $EW_{0}$'s ($\simeq 45-49$~\AA) and escape fractions ($\simeq 0.36-0.37$), are the ones that show weaker low-ionization ISM absorption ($EW_{0} \lesssim 1$\AA), narrower Ly$\alpha$ lines ($\rm FWHM \simeq 250-300$~km~s$^{-1}$) and smaller Ly$\alpha$ velocity offsets ($\simeq 30-80$~km~s$^{-1}$).

Overall, based on the properties of the Ly$\alpha$ line, such as the equivalent width (even noting that is likely underestimated), FWHM, $\Delta v~\rm (Ly\alpha)$, and escape fraction, as well as weak low-ionization ISM absorption lines, the BELLS GALLERY sources can be classified as LAEs, similar to those found in narrow-band imaging surveys. 
In addition, our galaxies show steep rest-frame UV slopes ($\overline{ \beta_{\rm UV}} = -1.7 \pm 0.2$), suggesting low dust content, and stellar metallicities of $Z_{*}/Z_{\odot} \simeq 0.2$ (see Section \ref{age}) similar to those inferred in other LAEs \citep[e.g.,][]{ono2010a, finkelstein2011a, nakajima2013, kojima2017}, but slightly lower than in the more evolved LBG population \citep[$\gtrsim 0.4 Z_{\odot}$, e.g.,][]{shapley2004, quider2009, quider2010, sommariva2012, marques2018, chisholm2019}. 
We note that the remaining BELLS GALLERY LAEs not observed with the GTC are likely to be even more extreme (larger $EW_{0}$'s~(Ly$\alpha$)) as they have fainter intrinsic UV luminosities \citep{shu2016b}, yet they show statistically larger observed Ly$\alpha$ fluxes within the $1^{\prime \prime}$-radius BOSS fibers (see Table 1 in \citealt{shu2016a}).

\subsection{AGN contribution to the UV and Ly$\alpha$ luminosities}\label{agn}

In this section we investigate the possible contribution of an Active Galactic Nucleus (AGN) to the UV and Ly$\alpha$ luminosities of our LAEs.

Starting with the UV continuum, the high S/N spectra of these LAEs exhibit both photospheric absorption lines and stellar wind features in the form of P-Cygni profiles (yellow and red lines in Figure \ref{fig:4}, respectively), clearly indicating that the UV luminosity is dominated by stellar emission rather than an AGN. 
Photospheric absorption lines are formed in the photospheres of young stars and are typically very faint ($EW_{0} < 1$~\AA; e.g., \citealt{prinja1990}). Because of their faintness, the detection of these stellar features requires high S/N continuum spectra, such as those recorded in other highly magnified galaxies \citep[e.g.,][]{pettini2000, quider2010, des2010, marques2018}, in composite spectra \citep[e.g.,][]{shapley2003, du2018} or in very deep spectroscopic observations of individual non-lensed sources \citep[$>20$h exposure time, e.g.,][]{marchi2019}. 
The strength of some of these features (e.g., Si~{\sc iii}~1417\AA, C~{\sc iii}~1426\AA, S~{\sc v}~1501\AA) are also known to be related with the stellar metallicity, such as they appear even weaker at lower stellar metallicities \citep[e.g.,][]{leitherer2001, rix2004, sommariva2012} like our LAEs ($Z/Z_{\odot} \simeq 0.2$, see Section \ref{age}). 
For that reason, we would not detect these stellar features if an AGN were present and its contribution to the UV continuum emission was significant (e.g., type I AGN). These LAEs also show narrow nebular lines ($<200$~km~s$^{-1}$ FWHM), including the Ly$\alpha$ resonant line ($\simeq 230-400$~km~s$^{-1}$) that is subject of scattering by neutral hydrogen. 
Furthermore, the lens modeling of these LAEs show that all of them present resolved structures \citep[see][and Section \ref{sec: lens}]{shu2016b, marques2017, cornachione2018}, removing most of the evidence for an AGN contribution to the UV luminosity.

\begin{figure}
\centering
  \includegraphics[width=0.4\textwidth]{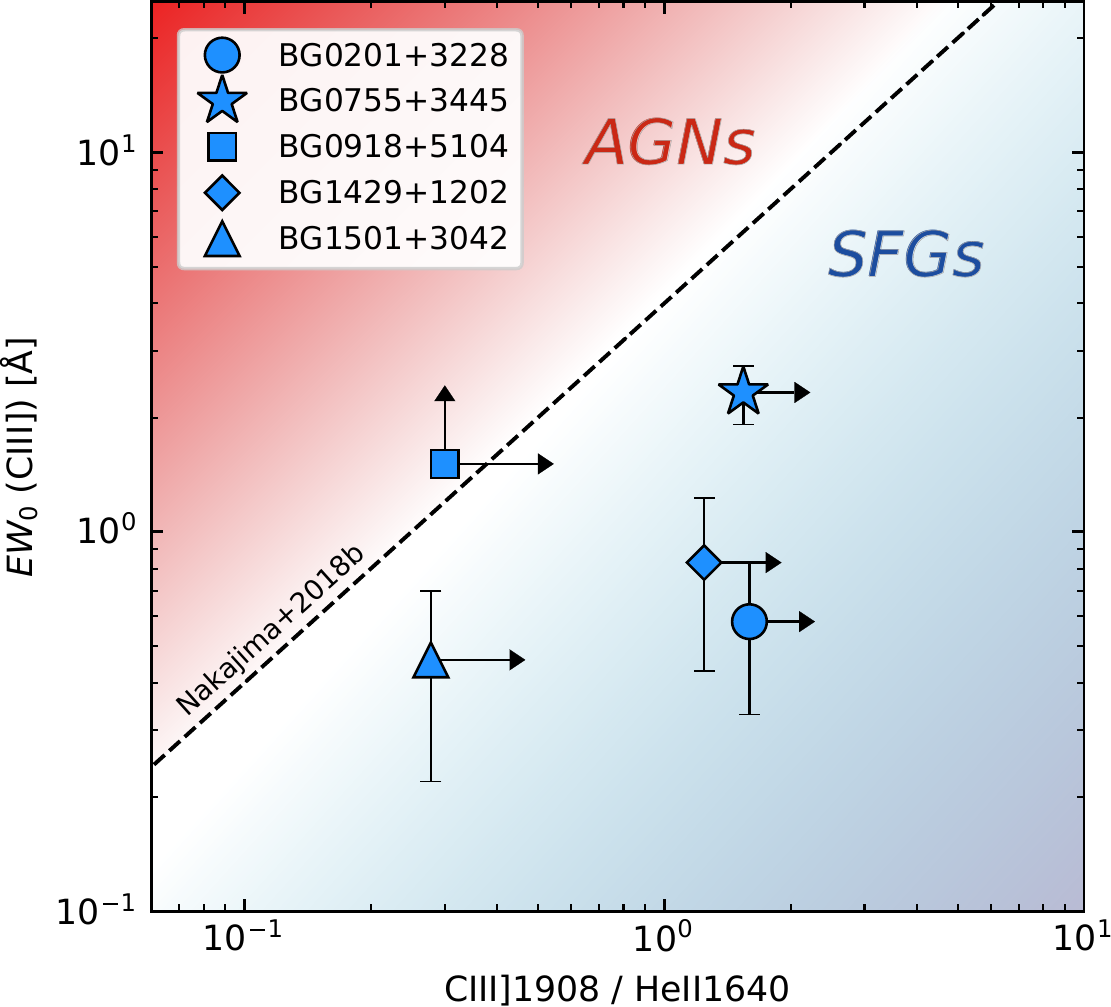}
    \caption{Positions of the BELLS GALLERY LAEs in the diagram of $EW_{0}$ (C {\sc iii}]) vs. C {\sc iii}]/He {\sc ii} proposed by \protect\cite{nakajima2017}. 
    This clearly suggests that the gas in these LAEs is excited by star formation and not by an AGN.}
    \label{fig:naka}
\end{figure}

On the contribution of a possible obscured AGN component to the Ly$\alpha$ luminosity, we explore the rest-frame UV emission-line diagnostics of \cite{nakajima2017} to identify the source of photoionization. These models were constructed covering a wide range of metallicities ($Z/Z_{\odot}$ from $10^{-4}$ to $5.0$), ionizing parameters ($\log (U)$ between $-3.5$ and $-0.5$) and electronic densities ($n=10 - 10^{5}$~cm$^{-3}$). Following \cite{nakajima2017}, we employ the UV line diagnostics using equivalent widths of the C~{\sc iii}]~1906,1908\AA{ }doublet and the line ratios of C~{\sc iii}] / He~{\sc ii} (Equation 2 in their work). Making use of the measurements of C~{\sc iii}] and He~{\sc ii} listed in Table \ref{tab:emission}, we find that the nebular emission in almost all LAEs is excited by star formation, rather than an AGN (see Figure \ref{fig:naka}). The exception is BG0918+5104, where their measurements of C~{\sc iii}] and the He~{\sc ii} upper limit are still consistent with the AGN photoionization. We note, however, that the [C~{\sc iii}]~1906\AA{ }nebular emission in BG0918+5104 is suppressed by the absorption of the Mg~{\sc ii}~2796,2803\AA{ }doublet from an intervening absorption system at $z = 1.312$ (see the lower panel of Figure \ref{fig:fs}), and that the spectral region encompassing the He~{\sc ii} line is highly affected by the [O~{\sc i}]~5577\AA{ }airglow emission, resulting in a very conservative upper limit of He~{\sc ii} (factor of $4-7$ larger than those derived in the remaining LAEs).

Therefore, we can conclude that both the UV and Ly$\alpha$ luminosities are due to star formation, rather than an AGN.

\begin{figure*}
  \centering
  \includegraphics[width=0.96\textwidth]{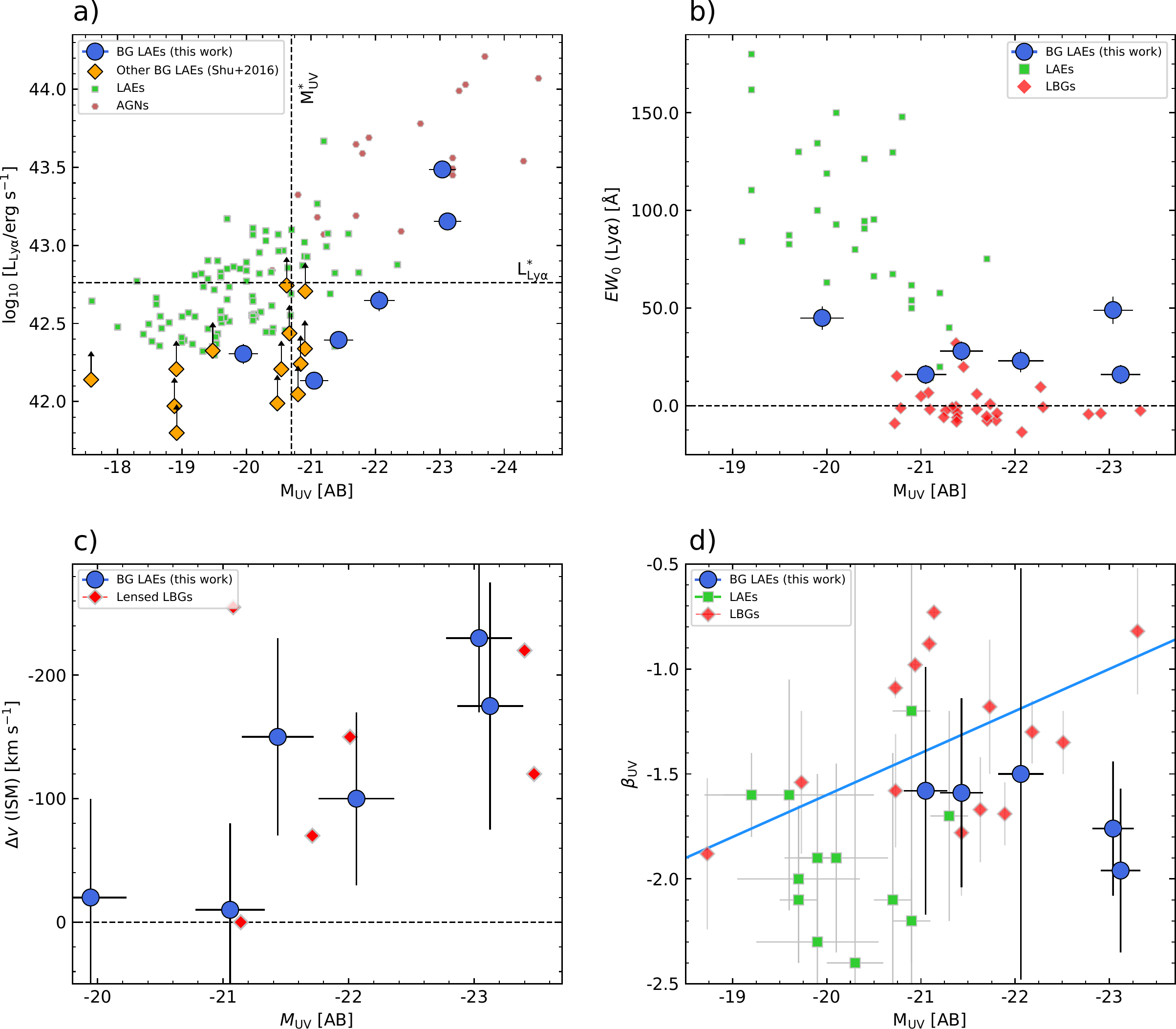}
  \caption{The relationship of rest-frame UV absolute magnitudes of our BELLS GALLERY LAEs (blue circles) on other properties: a) relation between rest-frame UV and Ly$\alpha$ luminosities. Horizontal and vertical dashed lines mark respectively the Ly$\alpha$ luminosity and UV absolute magnitude of typical $z\sim 2.5$ LAEs \protect\citep{sobral2018a} and LBGs \protect\citep{reddy2009}. Our LAEs span a wide range of luminosities, both in Ly$\alpha$ ($(0.2 - 5.4) \times L_{\rm Ly\alpha}^{*}$) and UV ($(0.5 - 10) \times L_{\rm UV}^{*}$). The remaining BELLS GALLERY LAEs imaged with {\it HST} \protect\citep{shu2016b} are also shown in orange. Other $z \sim 2 - 3$ LAEs and AGNs selected from dedicated narrow-band surveys with spectroscopic follow-up \protect\citep[][]{ouchi2008, zheng2016, sobral2018b} are shown in green squares and red circles, respectively.  
  b) Ly$\alpha$ equivalent widths as a function of $M_{\rm UV}$. Despite the large UV luminosities, our LAEs show prominent Ly$\alpha$ emission lines with $EW_{0}$ up to $\simeq 50$~\AA. Green and red symbols mark the measurements of other LAEs \protect\citep[][]{ouchi2008, sobral2018b} and UV-bright LBGs \protect\citep[][]{lee2013b};
  c) relation between the velocity offset of the ISM absorption lines relative the UV absolute magnitude. Red diamonds mark the measurements of other lensed LBGs \protect\citep[][]{pettini2000, quider2009, quider2010, des2010, patricio2016, marques2018}. Large outflowing velocities are noticeable for the most luminous LAEs and LBGs. This clearly suggests that the most luminous star-forming galaxies experience strong outflows caused by the large mechanical energy input from supernova explosions and stellar winds; and 
  d) UV slope ($\beta_{\rm UV}$) as a function of the intrinsic absolute magnitudes of our LAEs (blue). Other symbols mark these measurements in other $z=2-3$ LAEs \protect\citep[green;][]{sobral2018b, santos2019} and $z=2-4$ LBGs \protect\citep[orange;][]{shapley2003, bouwens2009, lee2011}. Blue line is the $\beta_{\rm UV} - M_{\rm UV}$ relation of \protect\cite{bouwens2009} for $z\sim 2.5$ LBGs. The luminous LAEs BG1429-1202 and BG1501+3042 ($M_{\rm UV} \sim -23$) show blue UV slopes despite their large luminosities and star formation rates.}
  \label{fig:last2}
\end{figure*}

\subsection{The nature of luminous LAEs}\label{sfr_agn}

Our LAEs present a wide range of Ly$\alpha$ luminosities, from $(1 - 31) \times 10^{42}$~erg~s$^{-1}$, which correspond to $(0.2 - 5.4) \times L_{\rm Ly\alpha}^{*}$, where $L_{\rm Ly \alpha}^{*}$ is the Ly$\alpha$ luminosity of typical $z \sim 2.5$ LAEs \citep[$L_{\rm Ly\alpha}^{*} = 5.8 \times 10^{42}$ erg s$^{-1}$;][]{sobral2018a}. 
In the UV continuum, the intrinsic absolute magnitudes of the LAEs are found to be in the range of $-20 \geq M_{\rm UV}~\rm (AB) \geq -23$. Considering the UV luminosity of typical $z \sim 2.3$ star-forming galaxies as $M_{\rm UV}^{*} = -20.70$ \citep{reddy2009}, these LAEs span the UV luminosity range from $0.5 \times L_{\rm UV}^{*}$ to $10 \times L_{\rm UV}^{*}$.

In Figure \ref{fig:last2}a we show the Ly$\alpha$ luminosities and rest-frame UV absolute magnitudes of the six BELLS GALLERY LAEs studied in this work. These values have been corrected for gravitational lensing and flux losses, and are summarized in Table \ref{all_laes}. In general, brighter $M_{\rm UV}$ objects have larger Ly$\alpha$ luminosities, as expected since both values increase with the SFRs.

We also add to Figure \ref{fig:last2} the remaining 13 BELLS GALLERY LAEs imaged with {\it HST}. To do so, we used the F606W magnification-corrected rest-frame UV magnitudes calculated from the best-fit source model presented in \cite{shu2016b}. 
Their Ly$\alpha$ luminosities were calculated using the fluxes measured from the $1^{\prime \prime}$-radius BOSS fibers \citep[see][]{shu2016a} after correction for the lensing magnification. We assume again similar lensing magnifications for the Ly$\alpha$ emission and the rest-frame UV continuum. 
We note, however, that these Ly$\alpha$ luminosities should be regarded as lower limits, since no corrections on the flux-losses from the BOSS fibers were applied. These galaxies present relatively fainter UV luminosities than those from the GTC survey, with UV absolute magnitudes ranging from $-17.6$ to $-21.0$~(AB).

In order to put the BELLS GALLERY LAEs into context, we compare their luminosities with other $z \sim 2 - 3$ LAEs selected from dedicated $1-4~\rm deg^{2}$ wide narrow-band surveys with spectroscopic follow-up \citep[][]{ouchi2008, zheng2016, sobral2018b}. 
From these surveys, a fraction of LAEs have been found to be  powered by AGNs (red symbols in panel a) of Figure \ref{fig:last2}), particularly in the bright end, by the detection of broad ($\rm FWHM > 1000$ km s$^{-1}$) emission lines (e.g., Ly$\alpha$, C~{\sc iv}, or C~{\sc iii]}), or slightly narrow high-ionization lines (such as N~{\sc v}, C~{\sc iv}, and He~{\sc ii}) with line ratios typically seen in AGNs \citep[e.g.,][]{hainline2011b}. 

Overall, our LAEs show statistically lower Ly$\alpha$ luminosities than those selected from narrow-band imaging surveys (except BG1429+1202 and BG1501+3042), yet relatively larger UV absolute magnitudes, likely a consequence of our selection strategy. The remaining BELLS GALLERY LAEs (in orange in panel a) of Figure \ref{fig:last2}) are likely to be more comparable to LAEs selected from narrow-band surveys (both in $L$~(Ly$\alpha$) and $M_{\rm UV}$) if the correction for slit-losses of the Ly$\alpha$ flux within the $1^{\prime \prime}$-radius BOSS fibers are similar to those of the GTC sample (a factor of $\sim 3-6$).

Recently, \cite{sobral2018b} showed that the fraction of AGN-dominated sources increases strongly with increasing Ly$\alpha$ and UV luminosities, reaching to about $100 \%$ for LAEs with luminosities above $2 \times L_{\rm Ly\alpha}^{*}$ or $L_{\rm UV}^{*}$ (i.e., $L_{\rm Ly\alpha} > 10^{43.3}$ erg s$^{-1}$ or $\rm M_{\rm UV} < -21.5$~AB). 
These authors argued that the dust formed in very luminous, purely star-forming galaxies prevents such galaxies being observed in the UV or Ly$\alpha$ if they are more luminous than this limit (see also: \citealt{konno2016}).  

However, as shown in Figure \ref{fig:last2} (panel a), there are several LAEs with luminosities larger than $2 \times L^{*}$ (in Ly$\alpha$ or UV), including three out of the six BELLS GALLERY LAEs observed with the GTC. 
Their spectra exhibit both photospheric absorption lines and stellar wind features, clearly indicates that the strong UV continuum and Ly$\alpha$ luminosity are due to star formation, rather than an AGN (see Section \ref{agn}).

Our results demonstrate that LAEs at $z \sim 2-3$ can be as luminous as $L_{\rm Ly\alpha} \simeq 10^{43.5}$~erg~s$^{-1}$ (or $\simeq 5 \times L_{\rm Ly\alpha}^{*}$) and $M_{\rm UV} \simeq -23$ AB (or $\simeq 10 \times L_{\rm UV}^{*}$) without invoking an AGN component, in contrast to the previous findings. 
It is worth noting that, while luminous LAEs at $z\sim 2-3$ appear scarce, as they have not been previously detected in these $1-4~\rm deg^{2}$ wide surveys \citep[][]{ouchi2008, zheng2016, sobral2018b}, a few UV-luminous ($M_{\rm UV} \sim -23$~AB) non-active LBGs at similar redshifts have been already discovered \citep[e.g.,][]{allam2007, lee2013b, lefevre2013, marques2018}. However, UV-luminous LBGs present on average a strongly suppressed Ly$\alpha$ line (see panel b of Figure \ref{fig:last2}) consistent with large neutral gas column densities \citep[$N$(H~{\sc i})>$10^{20}$~cm$^{-2}$,][]{des2010, marques2018}. Moreover, these luminous LBGs also appear more evolved than our luminous LAEs, as they show larger metallicities, stronger ISM absorption lines, and redder slopes suggesting large dust attenuation. Similar luminosities have been also found in very bright $z > 6$ UV- or Ly$\alpha$-selected star-forming galaxies \citep[e.g.,][]{ouchi2009, sobral2015, roberts2016, matthee2017, ono2018, shibuya2018, hashimoto2019, matsuoka2019}, although the contribution of a possible AGN needs to be investigated in more detail in future observations with the \textit{James Webb Space Telescope}.  

We also investigate the influence of such large luminosities and star-formation rates on the kinematics of the gas. 
In panel c) of Figure \ref{fig:last2} we compare the velocity offset of the ISM absorption lines relative to the systemic redshift and the absolute magnitudes of our LAEs. We also show these quantities measured in other lensed LBGs with damped Ly$\alpha$ absorption \citep[][]{pettini2000, des2010, quider2010, marques2018} or weak emission \citep[$EW_{0} < 20$~\AA:][]{quider2009, patricio2016}. These LBGs also present similar range of UV absolute magnitudes as our LAEs, from $-21$ to $-23.4$ (AB). In this case, both LAEs and LBGs appear to share the same global pattern. 
Large outflowing velocities of $\Delta v~\rm (ISM) \simeq -200$~km~s$^{-1}$ are more noticeable for the most luminous LAEs (BG1429+1202 and BG1501+3042) and LBGs (8~o'clock and HLock01-B galaxies: \citealt{des2010}, \citealt{marques2018}, respectively). 
For the less luminous ones, the ISM absorption lines have the minimum intensity very close to the systemic velocities.
This highly suggests that the most luminous star-forming galaxies, the ones that show larger star formation rates, experience strong outflows caused by the mechanical energy input from supernova explosions and stellar winds. 

More interestingly, the most luminous LAEs in our sample, BG1429+1202 and BG1501+3042 (with $M_{\rm UV}\sim -23$), also present bluer $\beta_{\rm UV}$ slopes than the fainter ones (panel d in Figure \ref{fig:last2}), in contrast to the general idea that $\beta_{\rm UV}$ is much redder at higher luminosities and SFRs \citep[e.g.,][]{bouwens2009, lee2011, hashimoto2017, oyarzun2017}. In addition, other spectral properties of these luminous LAEs are broadly similar to those found in much fainter LAEs: e.g., large $f_{\rm esc}~\rm (Ly\alpha)$, weak ISM absorption lines and low stellar metallicity; all of them thought to be closely related to an early evolutionary state of a galaxy. 

It is possible therefore that these luminous LAEs are experiencing a very recent and massive burst of star formation, so that they have only had time to form small amounts of dust. 
Recent close-by merger events or accretion of gas may have triggered the large luminosities and SFRs of these galaxies. For example, BG1501+3042 is composed by two UV-bright components separated by only $\simeq 4$~kpc in projection (see Figure \ref{fig:bg1501_lens}), highly suggesting a major merger event. Similarly, the reconstructed source-plane image of BG0201+3228 also reveals three UV-bright components separated by only $\sim 1.2$~kpc \citep{shu2016b, cornachione2018}. Additionally, 
the extreme feedback associated with the large SFRs may produce channels through the ISM where Ly$\alpha$ photons could escape more efficiently. A clumpy ISM could explain the relatively weak ISM absorption lines (and neutral gas covering fractions) and the large Ly$\alpha$ escape fractions in these luminous LAEs.

Regardless the mechanisms responsible for the large luminosities (recent episodes of merger, interaction, or accretion of gas), their vigorous SFRs will likely change dramatically their chemical and dust composition in a very short timescale, converting them in more evolved LBGs. Such a short timescale could in principle explain the rarity of these UV/Ly$\alpha$ luminous systems.

\section{Summary and Conclusions} \label{sec:conclusion}

This work presents the results of a GTC and WHT imaging and spectroscopic follow-up of six BELLS GALLERY lensed LAEs. They were chosen to have large image separations and flux densities in optical bands in order to perform high S/N studies of their physical properties. These LAEs, with redshifts between $2.36$ and $2.82$, are lensed by massive early type galaxies at $z \simeq 0.39 - 0.72$ providing total magnifications of $\simeq 3 - 18$. From the analysis of these data we obtain the following results:

\begin{enumerate}

\item We spectroscopically confirmed the lensing nature of these systems through the detection of emission/absorption spectral features in several lensed images of each LAE. One of these, BG1501+3042 at $z=2.645$, is a new confirmed lensed LAE with three detected lensed images. This system is composed by a pair interacting/merging galaxies magnified by a factor of $\simeq 3.2$. 
In addition, we measured the systemic redshifts of all LAEs using nebular emission from H~{\sc ii} regions, stellar absorption from the photospheres of massive stars, and fine-structure emission lines. 

\item These LAEs show similar properties to those selected from narrow-band imaging surveys, even though they have been selected through different methods. We measure Ly$\alpha$ rest-frame equivalent widths ranging from 16 \AA{ }to 50 \AA{ }and Ly$\alpha$ escape fractions ranging from 10\% to 40\% assuming case-B recombination. These measurements are likely underestimated due to larger slit-losses of the Ly$\alpha$ flux relative to the UV continuum. These LAEs show weaker low-ionization ISM lines when compared to the more evolved LBG population. In addition, we find that for almost all LAEs their N~{\sc v} and C~{\sc iv} stellar P-Cygni profiles are well fitted by {\sc Starburst99} stellar synthesis models with metallicity of $Z/Z_{\odot} = 0.2$.

\item We measure dust-corrected star formation rates up to $\simeq 180$ $M_{\odot}$ yr$^{-1}$, yet these LAEs appear relatively blue ($\beta_{\rm UV} \simeq -1.5$ to $-2.0$) showing evidence of little dust attenuation, $E(B-V) = 0.10-0.20$. They present rest-frame UV and Ly$\alpha$ intrinsic luminosities in the range of $-20 <M_{\rm UV}<-23 $ and $L_{\rm Ly\alpha} = (1 - 30) \times 10^{42}$~erg~s$^{-1}$, respectively. None of them show any evidence for AGN activity. Even knowing that luminous LAEs ($> 2 \times L^{*}$) are likely rare, they exist also at $z =2-3$ and can show luminosities at least up to $L_{\rm Ly\alpha} \simeq 10^{43.5}$~erg~s$^{-1}$ and $M_{\rm UV} \simeq -23$~AB, without
invoking to an AGN component, in contrast with previous findings. 

\item We show that $f_{\rm esc}$~(Ly$\alpha$) increases with increasing $EW_{0}$~(Ly$\alpha$). We find that LAEs with lower $EW_{0}$~(Ly$\alpha$) and $f_{\rm esc}$~(Ly$\alpha$) present broader Ly$\alpha$ line profiles with their peak more redshifted relative to their systemic redshifts, by $\simeq 200 - 300$ km s$^{-1}$, consistent with Ly$\alpha$ photons being scattered by receding outflowing gas. On the other hand, for LAEs with larger $EW_{0}$~(Ly$\alpha$) and $f_{\rm esc}$~(Ly$\alpha$), the Ly$\alpha$ line peaks close to (or consistent with) the systemic redshift. The Ly$\alpha$ line in these ones also present narrower profiles, suggestive of less scatter from low H~{\sc i} column densities that favours the escape of Ly$\alpha$ photons.

\item On the kinematics, the centroids of the ISM absorption lines in some of these LAEs appear blueshifted up to $200$ km s$^{-1}$ relative to their systemic redshift, indicative of large galaxy-scale outflows due to radiation pressure of the most massive and luminous stars. These LAEs are the ones that present larger UV and Ly$\alpha$ luminosities, and therefore, larger star formation rates. On the other hand, for the less luminous LAEs, the ISM absorption lines have the minimum intensity close to the systemic velocity.

\item We identify a trend for $EW_{0}$~(Ly$\alpha$) to increase with decreasing $EW_{0}$~(low-ISM).
Low-ionization ISM lines show residual intensities of about $I/I_{0} \simeq 0.3 -0.7$ indicating partial coverage of the interstellar gas. Assuming the optically thick regime and using the low-ionization ISM line Si~{\sc ii}~1260\AA, we measure dust-corrected covering fractions of neutral gas $C_{f}$~(H~{\sc i}) between $\sim 0.68-0.96$. 

\end{enumerate} 

Overall, we have shown that LAEs can present a wide range of luminosities and are not restricted to the faint end only, as previously thought. 
Interestingly, the most luminous ones also present steeper UV slopes. We speculate that these luminous LAEs may be particular cases of very young starburst galaxies that have only had time to form small amounts of dust. If so, this phase should last only a few Myr, as their large SFRs will likely enrich the chemical and dust composition, converting these LAEs in more evolved LBGs. 
Such a short timescale could in principle explain the very low number density of these UV and Ly$\alpha$ luminous systems.

\section*{Acknowledgements}

The authors thank the anonymous referee for useful comments. Based on observations made with the Gran Telescopio Canarias (GTC) and with the William Herschel Telescope (WHT), installed in the Spanish Observatorio del Roque de los Muchachos of the Instituto de Astrof\'{i}sica de Canarias, in the island of La Palma. We thank the GTC and WHT staff for their help with the observations. 
R.M.C. would like to thank Claudio Dalla~Vecchia, Pablo P\'{e}rez-Gonz\'{a}lez, and Rosa Gonz\'{a}lez-Delgado for their comments on the first version of the paper.
R.M.C. acknowledges Fundaci\'{o}n La Caixa for the financial support received in the form of a Ph.D. contract. This research has been funded by the Spanish State Research Agency (AEI) Projects No. ESP2017-83197 and No. MDM-2017-0737 Unidad de Excelencia ``Mar\'{i}a de Maeztu'' - Centro de Astrobiolog\'{i}a (CSIC-INTA). 
R.M.C., I.P.F., L.C., J.A.M., C.J.A, F.P. and R.S. acknowledge support from the Spanish Ministerio de Ciencia, Innovaci\'{o}n y Universidades (MICINN) under grant numbers ESP2015-65597-C4-4-R, and ESP2017-86852-C4-2-R. 
M.A.C acknowledges support by the National Science Foundation under grant AST-1614018. A.D.M.D. thanks FAPESP for financial support.

Funding for the Sloan Digital Sky Survey IV has been provided by the Alfred P. Sloan Foundation, the U.S. Department of Energy Office of Science, and the Participating Institutions. SDSS-IV acknowledges support and resources from the Center for High-Performance Computing at the University of Utah. The SDSS web site is www.sdss.org. SDSS-IV is managed by the Astrophysical Research Consortium for the Participating Institutions of the SDSS Collaboration including the Brazilian Participation Group, the Carnegie Institution for Science, Carnegie Mellon University, the Chilean Participation Group, the French Participation Group, Harvard-Smithsonian Center for Astrophysics, Instituto de Astrof\'isica de Canarias, The Johns Hopkins University, Kavli Institute for the Physics and Mathematics of the Universe (IPMU) / University of Tokyo, the Korean Participation Group, Lawrence Berkeley National Laboratory, Leibniz Institut f\"ur Astrophysik Potsdam (AIP), Max-Planck-Institut f\"ur Astronomie (MPIA Heidelberg), Max-Planck-Institut f\"ur Astrophysik (MPA Garching), Max-Planck-Institut f\"ur Extraterrestrische Physik (MPE), National Astronomical Observatories of China, New Mexico State University, New York University, University of Notre Dame, Observat\'ario Nacional / MCTI, The Ohio State University, 
Pennsylvania State University, Shanghai Astronomical Observatory, United Kingdom Participation Group, Universidad Nacional Aut\'onoma de M\'exico, University of Arizona, University of Colorado Boulder, University of Oxford, University of Portsmouth, 
University of Utah, University of Virginia, University of Washington, University of Wisconsin, Vanderbilt University, and Yale University.

\bibliographystyle{mnras}
\input{ms_BELLS_GALLERY_GTC.bbl}

\begin{appendix}

\section{Lower-z Intervening Metal-Line Systems}\label{low-z}

In addition to the absorption lines associated with the interstellar medium gas and stars of each lensed LAE, we also detect other absorption lines from intervening metal-systems at lower redshifts. 
In total we find 5 intervening metal-systems at redshifts between $1.31$ and $2.70$. None of these systems 
is associated to the LAEs or their corresponding lens, but originates in intervening systems along the line of sight to the LAEs.
Table \ref{tab:lowz} lists the identified intervening metal-absorption lines for each lensed LAE with their redshifts and rest-frame equivalent widths.

\begin{table}
\begin{center}
\caption{Intervening Metal-Absorption Line Systems.}\label{tab:lowz}
 \begin{tabular}{llcc}
\hline  \hline 
$\lambda_{\rm obs}$ (\AA)  & Line & $z_{\rm abs}$ & $EW_{0}$ (\AA) \\  
\hline
\multicolumn{4}{c}{BG0201+3228 $-$ System 1: $z_{\rm abs} = 2.705$}\\ 
\hline 
5164.45 & Si~{\sc iv} 1393.75 & 2.7055 & 0.47 \\
5197.65 & Si~{\sc iv} 1402.77 & 2.7050 & 0.45 \\
5734.79$^{\rm a}$ & C~{\sc iv}  1548.19 & 2.7042$^{\rm a}$ & 0.38$^{\rm a}$ \\ 
5734.79$^{\rm a}$ & C~{\sc iv}  1550.77 & 2.7040$^{\rm a}$ & 0.57$^{\rm a}$ \\ 
6872.02$^{\rm b}$ & Al~{\sc iii} 1854.71 & 2.7052$^{\rm b}$ & 0.89$^{\rm b}$ \\
6900.53 & Al~{\sc iii} 1862.79 & 2.7044 & 0.94 \\
\hline 
\multicolumn{4}{c}{BG0918+5104 $-$ System 1: $z_{\rm abs} = 1.3121$}\\ 
\hline
5419.80 & Fe~{\sc ii} 2344.21 & 1.3120 & 0.47 \\
5490.11 & Fe~{\sc ii} 2374.46 & 1.3121 & 0.09 \\
5508.68 & Fe~{\sc ii} 2382.76 & 1.3119 & 0.72 \\
6011.78 & Fe~{\sc ii} 2600.17 & 1.3121 & 0.65 \\
6465.10 & Mg~{\sc ii} 2796.35 & 1.3120 & 0.86 \\
6482.65$^{\rm c}$ & Mg~{\sc ii} 2803.53 & 1.3123$^{\rm c}$ & 0.29$^{\rm c}$ \\
\hline 
\multicolumn{4}{c}{BG1429+1202 $-$ System 1: $z_{\rm abs} = 2.1804$}\\ 
\hline
4855.53 & Si~{\sc ii} 1526.70 & 2.1804 & 0.75 \\
4923.25 & C~{\sc iv}  1548.19 & 2.1800 & 0.75 \\
4931.90 & C~{\sc iv}  1550.77 & 2.1803 & 0.38 \\
5115.58 & Fe~{\sc ii} 1608.45 & 2.1804 & 0.49 \\
5313.99 & Al~{\sc ii} 1670.78 & 2.1805 & 0.88 \\
7455.35 & Fe~{\sc ii} 2344.21 & 2.1803 & 0.70 \\
7551.81 & Fe~{\sc ii} 2374.46 & 2.1804 & 0.50 \\
7578.35 & Fe~{\sc ii} 2382.76 & 2.1805 & 1.04 \\
\hline 
\multicolumn{4}{c}{BG1429+1202 $-$ System 2: $z_{\rm abs} = 2.3472$}\\ 
\hline
5182.16 & C~{\sc iv} 1548.19 & 2.3472 & 1.37 \\
5190.61 & C~{\sc iv}  1550.70 & 2.3473 & 0.92 \\
\hline 
\multicolumn{4}{c}{BG1502+3042 $-$ System 1: $z_{\rm abs} = 2.269$}\\ 
\hline
4556.67 & Si~{\sc iv} 1393.75 & 2.2694 & 1.67 \\
4584.17$^{\rm d}$ & Si~{\sc iv} 1402.77 & 2.2680$^{\rm d}$ & 1.22$^{\rm d}$ \\
5064.80$^{\rm e}$ & C~{\sc iv} 1548.19 & 2.2687$^{\rm e}$ & 1.84$^{\rm e}$ \\
5064.80$^{\rm e}$ & C~{\sc iv}  1550.70 & 2.2687$^{\rm e}$ & 1.84$^{\rm e}$ \\
\hline
\end{tabular}
\end{center}
\textbf{Notes.} 
$^{\rm a}$ Blended 
with the stellar photospheric absorption S~{\sc V} $\lambda$1501 of BG0201+3228.
$^{\rm b}$ This intervening absorption line falls in the region of a telluric absorption line (B-band).
$^{\rm c}$ Blended 
with the nebular emission C~{\sc iii}] $\lambda$1906 of BG0918+5104.
$^{\rm d}$ Blended 
with the absorption from Si~{\sc ii} $\lambda$1260 of BG1502+3042.
$^{\rm e}$ The measured values refer to the C~{\sc iv} $\lambda$1548,1550 doublet given the 
low spectral resolution of the GTC R1000B grism. 
\end{table}

\end{appendix}
\label{lastpage}
\end{document}